\documentclass[journal,10pt]{IEEEtran}
\usepackage{amsthm}
\usepackage{amsmath}
\usepackage{graphicx}
\usepackage{algorithm}
\usepackage{algorithmicx}
\usepackage{algpseudocode}
\usepackage{subfigure}
\usepackage[]{mcode}

\ifCLASSINFOpdf
\else
\fi
\usepackage{pdfpages}

\begin{document}
%
\title{One-Sided Box Filter for Edge Preserving Image Smoothing}
%
%
%

\author{Yuanhao~Gong\\College of Electronics and Information Engineering, Shenzhen University, China, gong@szu.edu.cn
	\thanks{Yuanhao Gong is with College of Electronics and Information Engineering, Shenzhen University, China. This work was supported in part by the National Natural Science Foundation of China under Grant 61907031.}
	}

%
%

\markboth{One-Sided Box Filter}%
{Yuanhao: One-Sided Box Filter}
%



\maketitle

\begin{abstract}
Image smoothing is a fundamental task in signal processing. For such task, box filter is well-known. However, box filter can not keep some features of the signal, such as edges, corners and the jump in the step function. In this paper, we present a one-sided box filter that can smooth the signal but keep the discontinuous features in the signal. More specifically, we perform box filter on eight one-sided windows, leading to a one-sided box filter that can preserve corners and edges. Our filter inherits the constant $O(1)$ computational complexity of the original box filter with respect to the window size and also the linear $O(N)$ computational complexity with respect to the total number of samples. We performance several experiments to show the efficiency and effectiveness of this filter. We further compare our filter with other the-state-of-the-art edge preserving methods. Our filter can be deployed in a large range of applications where the classical box filter is adopted.
\end{abstract}

\begin{IEEEkeywords}
one-sided, Sub-window Regression, Gaussian Filter, Box Filter
\end{IEEEkeywords}

%
\IEEEpeerreviewmaketitle

\section{introduction}
%
%
%
%
\IEEEPARstart{I}mage smoothing has been playing an important role in various image processing tasks. It is widely used in image denoising~\cite{gong2013a,gong:gdp}, compression~\cite{Burt:1983}, interpolation~\cite{Zhang:2006}, segmentation~\cite{Gong2012}, deconvolution~\cite{gong:gdp}, enhancement~\cite{Gong:2014a}, etc. In these tasks, smoothing can remove the noise and unnecessary details of the image. Therefore, successfully performing image smoothing is the key to accomplish these tasks.

The two dimensional image smoothing process can be generally written as 
\begin{equation}
U(x,y)=\Phi(I({\cal N}(x,y)))\,,
\end{equation} where $U(x,y)$ and $I(x,y)$ denote the smoothed image and the observed input image, respectively. $\Phi$ is a function than performs the transformation between $U$ and $I$. The $(x,y)\in\Omega$ denotes the spatial coordinate and $\Omega$ is the imaging domain. ${\cal N}(x,y)$ denotes the neighborhood or support region for the location $(x,y)$.

Various approaches have been developed for the image smoothing task. One way of categorizing these smoothing methods is by ${\cal N}$. If ${\cal N}=\Omega$, the method is called a global method; otherwise, it is called a local method. Global methods (for example, moving least square method (MLS)~\cite{fleishman2005robust,gong2013a}) can exchange the global spatial information thanks to the long range interaction. Because of the global support region, they usually require to construct a very large matrix. On the other hand, local methods (such as bilateral filter~\cite{Tomasi:1998,Zhang:2012,Pan:2014,gong2009symmetry,gong:phd}) avoid such large matrix. These local methods only relay on local neighbor samples. Therefore, their long range information exchange would take a large number of iterations because the local window limits the speed of information propagation.

\begin{figure}[!t]
	\centering
		\subfigure[One-Sided Box Filter with iteration 0, 2, 10 and 30. (window radius $r=2$)]{
		\includegraphics[width=\linewidth]{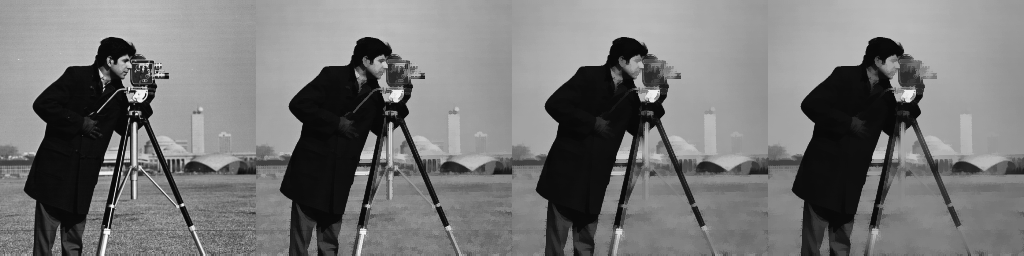}}
	\subfigure[Classical Box Filter with iteration 0, 2, 10 and 30. (window radius $r=2$)]{
		\includegraphics[width=\linewidth]{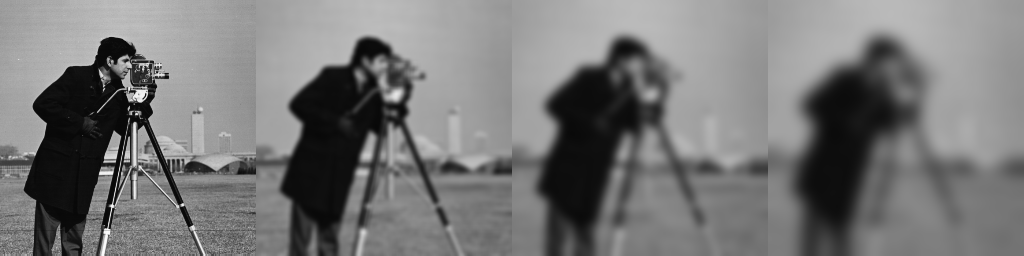}
	}
	
	\caption{Comparison of the presented one-sided box filter and the classical box filter with different iteration number. The window radius $r=2$ for both methods. One-sided box filter can preserve edges and corners. But the original box filter can NOT preserve edges and corners. Details are in later sections.}
	\label{fig:subBoxFilter}
\end{figure}

From edge preserving perspective, we can categorize these smoothing methods into edge-preserving methods and not edge-preserving methods. If one method smooths out the large gradient, it is not edge preserving. Such methods include the classical isotropic diffusion filters like Gaussian filter and box filter. In contrast, one smoothing method can be called an edge preserving method if it removes the small gradient of the input image but preserves the large gradient. Typical edge preserving methods include bilateral filter~\cite{Tomasi:1998} and side window filters~\cite{Yin2019}. 

\begin{figure}[!tb]
	\centering
	\subfigure[full window]{\includegraphics[width=0.23\linewidth]{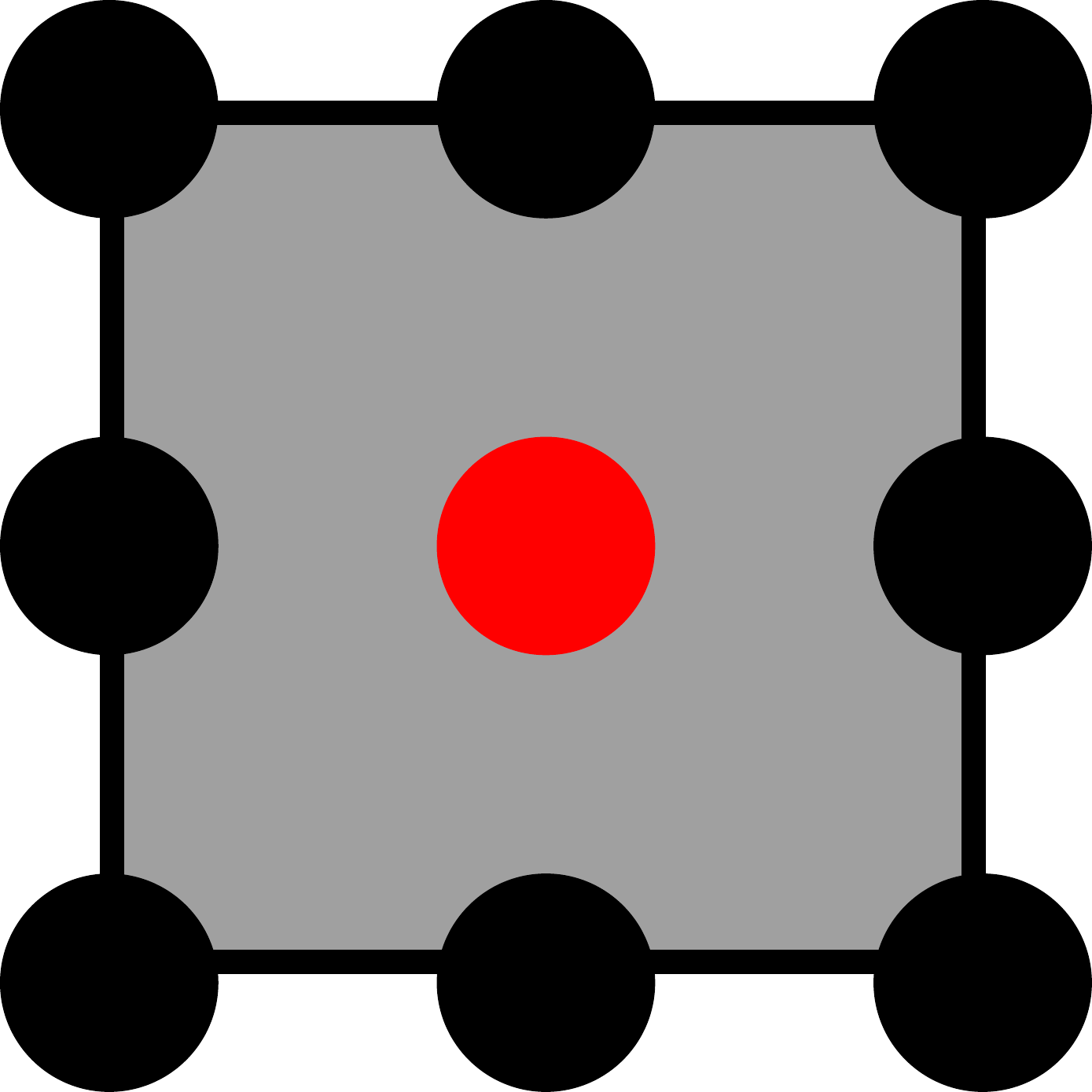}}
	\subfigure[anisotropic]{\includegraphics[width=0.23\linewidth]{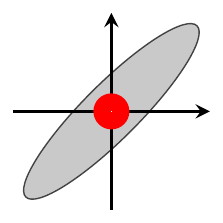}}
	\subfigure[adaptive scale]{\includegraphics[width=0.25\linewidth]{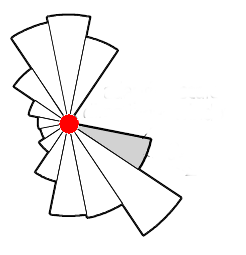}}
	\subfigure[cross  window]{\includegraphics[width=0.23\linewidth]{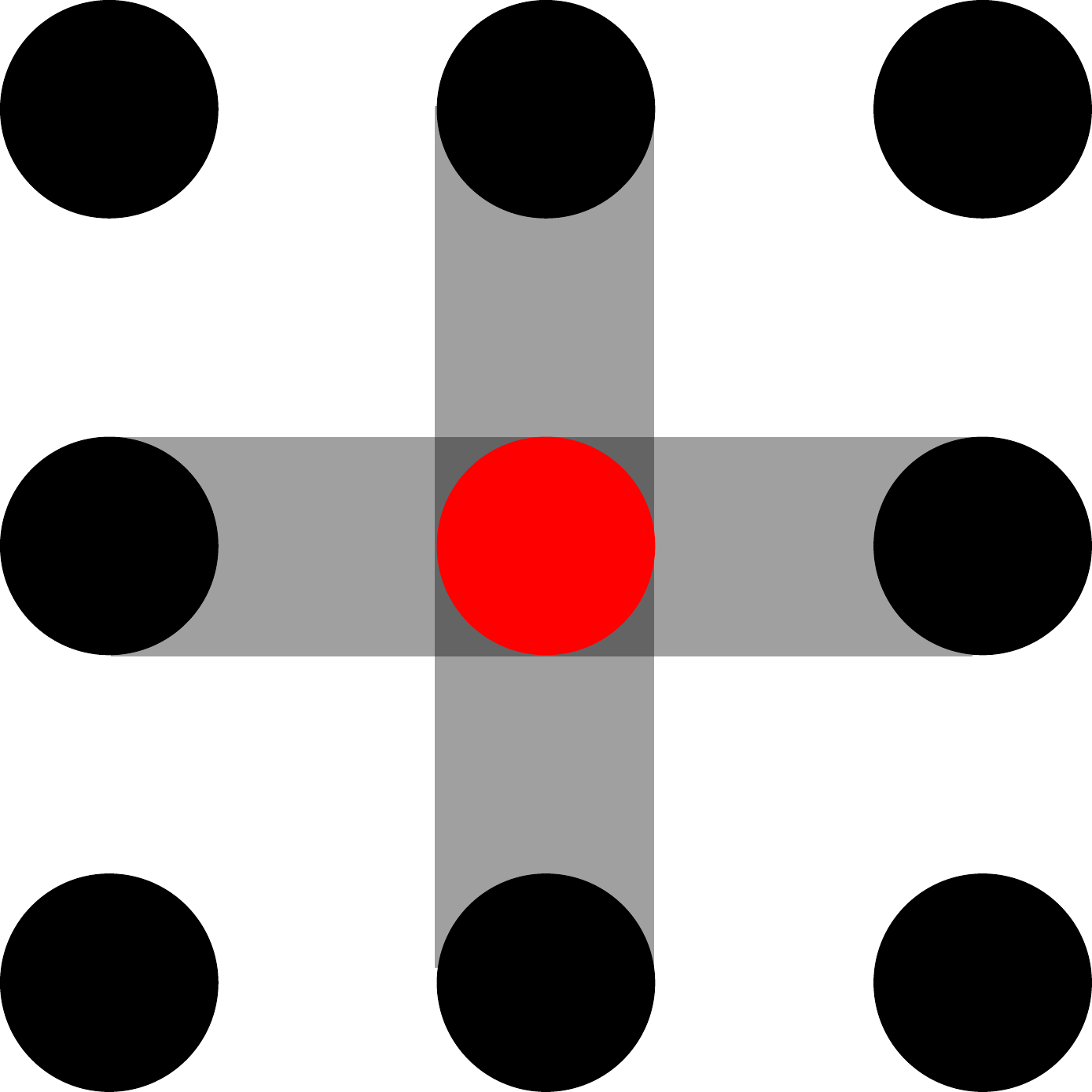}}
	\caption{Four typical support regions ${\cal N}$. The pixel value at the red dot position is estimated by the shaded area in each case. (a) has fixed support regions. (b), (c) and (d) have image content adaptive regions.}
	\label{fig:illus}
\end{figure}

Thanks to the computational efficiency and effectiveness, the local edge preserving methods are getting popular in the past few years. These filters estimate the smoothed result by performing a regression on a local neighborhood window ${\cal N}(x,y)=\{(u,v)||u-x|\leq r,|v-y|\leq r\}$, where $r$ is the local window radius. In the discrete case, $r$ is an integer. This local window has width $2r+1$ and contains $(2r+1)^2$ pixels. Such square window is illustrated in Fig.~\ref{fig:illus} (a). 

The local methods usually take following expression
\begin{equation}
U(x,y)=\int\limits_{u=-r}^{r}\int\limits_{v=-r}^{r}k(u,v)I(x-u,y-v)\mathrm{d}u\mathrm{d}v\,,
\end{equation} where $k(u,v)\geq 0$ and $\int\int k(u,v)\mathrm{d}u\mathrm{d}v=1$. The $k(u,v)$ is called the smooth kernel or convolution kernel. In such methods, the kernel $k$ plays a fundamental role. In fact, for local methods, $k$ uniquely determines the transfer function $\Phi$. In the rest of this paper, we use $\Phi$ to indicate the general transfer function and use $k$ only for kernels in local methods.

According to the transform function $\Phi$ and the supported regions ${\cal N}$, we can classify the smoothing methods into four classes. The first one is the content independent $\Phi$ on content independent ${\cal N}$, which will be discussed in Section~\ref{sec:first}. The second class is content adaptive $\Phi$ on content independent ${\cal N}$, which will be discussed in Section~\ref{sec:second}. The third class is content adaptive $\Phi$ on content adaptive ${\cal N}$, , which will be discussed in Section~\ref{sec:three}. The last one is content independent $\Phi$ on content adaptive ${\cal N}$, which will be discussed in Section~\ref{sec:four}.

\subsection{Content Independent $\Phi$ on Content Independent ${\cal N}$}
\label{sec:first}
Traditional methods tried various weight functions for $\Phi$ because it, indeed, is important in filter designing. They usually take the $(2r+1)\times(2r+1)$ square window as shown in Fig.~\ref{fig:illus}(a) and perform $\Phi$ on this local window. One typical example is to use the constant function as weight in the local window, which leads to the classical box filter. The weight function in box filter is independent of the image content, which reduces its computational cost. However, this box filter can not preserve image edges because it is an isotropic diffusion process. 

\subsection{Content Adaptive $\Phi$ on Content Independent ${\cal N}$}
\label{sec:second}
To achieve edge-preserving, many different $\Phi$ were introduced. One strategy is to use a weight function whose value not only relies on spatial coordinates but also the image intensity, such as bilateral filter~\cite{Tomasi:1998,gong2009symmetry}. Another way to preserve edges is to imposing gradient domain regression, such as trilateral filter~\cite{Choudhury:2005}, guided filter~\cite{he2010guided} and weighted guided filter~\cite{zhengguo:2015}. Besides the zero order (constant intensity) and first order information (gradient), it is also possible to involve higher order information on this local window, such as kernel regression~\cite{takeda:2007} and curvature information~\cite{gong2013a,Gong2012,Gong:2014a,gong:cf,GONG2019329}. Although the weight function is adaptive to the image content, the support region of these methods is still the full local window. This case is illustrated in Fig.~\ref{fig:illus} (a). 

However, most of these filters suffer from generating halo artifacts or the heavy computation cost~\cite{he2010guided,zhengguo:2015,takeda:2007}. The halo artifacts come from the fact that the imposed prior information might be invalid for the input data. The heavy computation comes from the fact that the weight function is spatially varying and depends on the image content. In other word, the weight function has to be evaluated for each local window. Such computation becomes challenge for high resolution images. 

\subsection{Content Adaptive $\Phi$ on Content Adaptive ${\cal N}$}
\label{sec:three}
To obtain edge preserving, instead of focusing on $\Phi$, some methods focus on the shapes of ${\cal N}$. One typical example is the famous adaptive scale directional window~\cite{Foi:2007}, as shown in Fig.~\ref{fig:illus} (c). Although this adaptive window can give better results, it is not easy to find the orientation and scale for each pixel because the window shapes depend on the image content. Another example is to adopt the cross base~\cite{CLMF}, as shown in Fig.~\ref{fig:illus}(d). But the four arm lengths of the cross spatially change depending on the image content. Even though both can get better result, they are suffering from the heavy computation because the adaptive shapes ${\cal N}$ have to be estimated for each individual pixel.  

Recently, curvature filters are proposed to minimize Gaussian or mean curvatures in the local window and they perform on several $3\times3$ half windows~\cite{gong:cf}. These filters are several orders of magnitude faster than traditional geometry flows because the differential geometry is implicitly encoded. But these methods are difficult to be extended onto a larger window and for higher dimensional data, such as video and biomedical volume data. 

\subsection{Content Independent $\Phi$ on Content Adaptive ${\cal N}$}
\label{sec:four}

\begin{figure}
	\centering
	\includegraphics[width=0.9\linewidth]{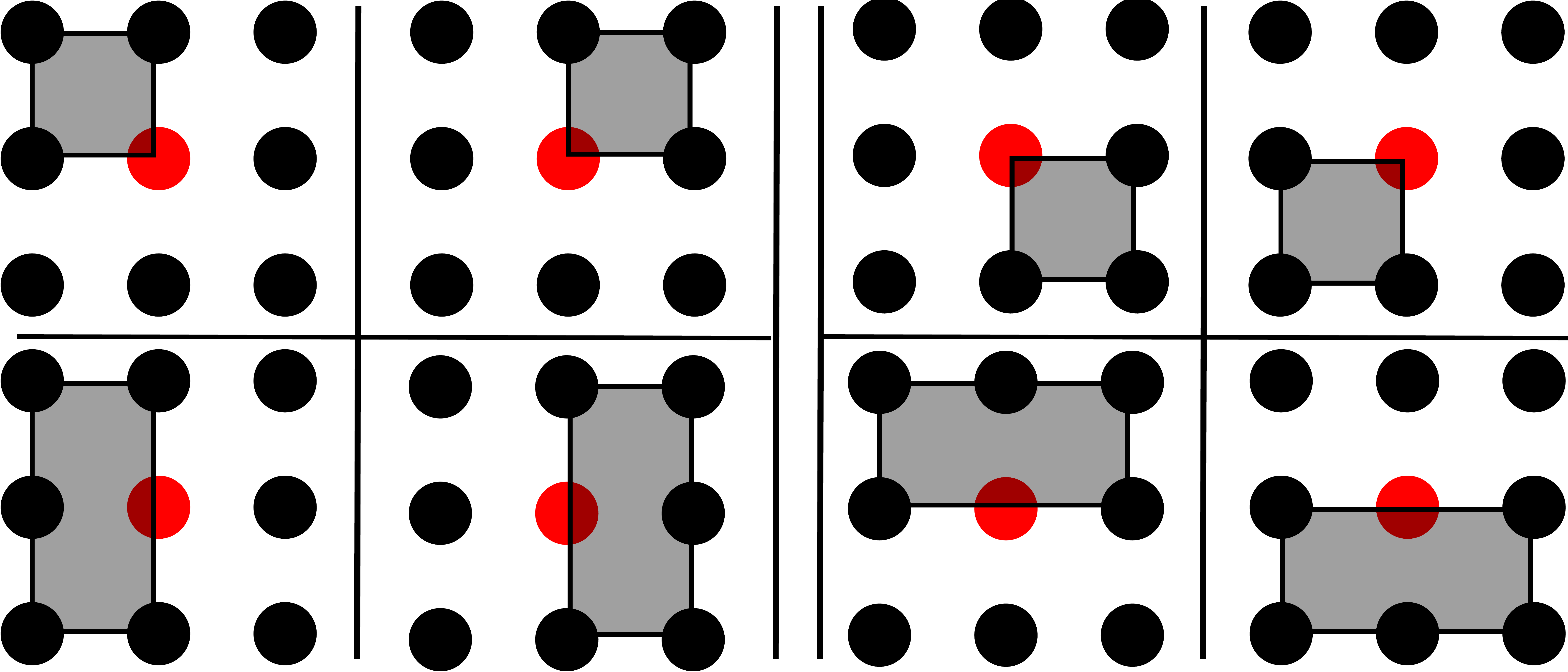}
	\caption{Eight possible support regions for the red central location. Although eight regions are computed, only one is selected for the center location. Be aware that the overlapped regions can be used to accelerate the computation.}
	\label{fig:kernels8}
\end{figure}

To the best of our knowledge, there is no work that focuses on content independent $\Phi$ and content adaptive ${\cal N}$. In this paper, we propose such methods. More specifically, we propose to perform $\Phi$ on several one sided windows with arbitrary radius $r$, as shown in Fig.~\ref{fig:kernels8}, and choose one of them as the result. The reason is that the central pixel can only belong to one of these eight regions. In the first row of Fig.~\ref{fig:kernels8}, we assume the central pixel has a corner geometry. In the second row of Fig.~\ref{fig:kernels8}, we assume the central pixel is on an edge. When the central pixel is in a flat region, choosing which case is not important. But when the central pixel is at a corner or an edge, it is important to choose the correct support region. 

\begin{figure}[!htb]
	\centering
	\subfigure[Four ignored triangular half windows]{
		\includegraphics[width=0.9\linewidth]{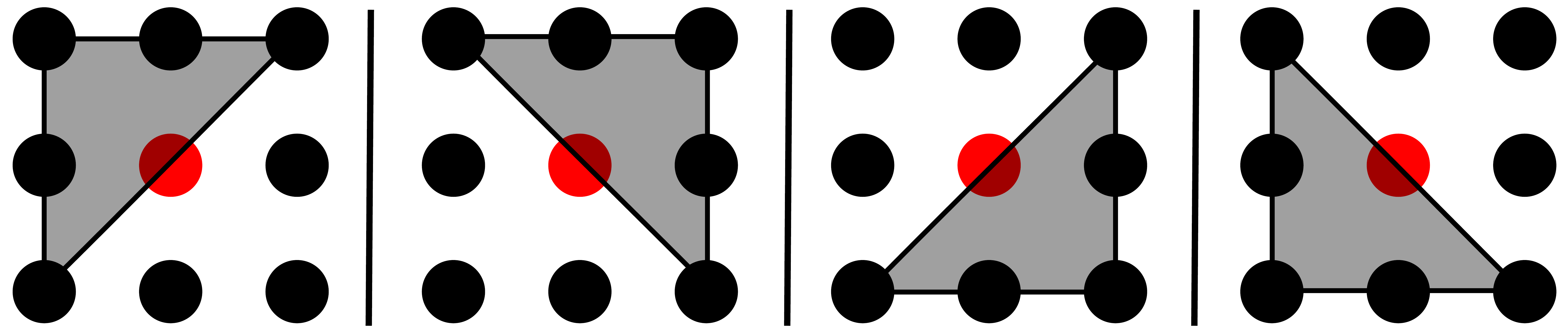}}
	\subfigure[Approximating diagonal regions by our quarter rectangular regions]{
		\includegraphics[width=0.9\linewidth]{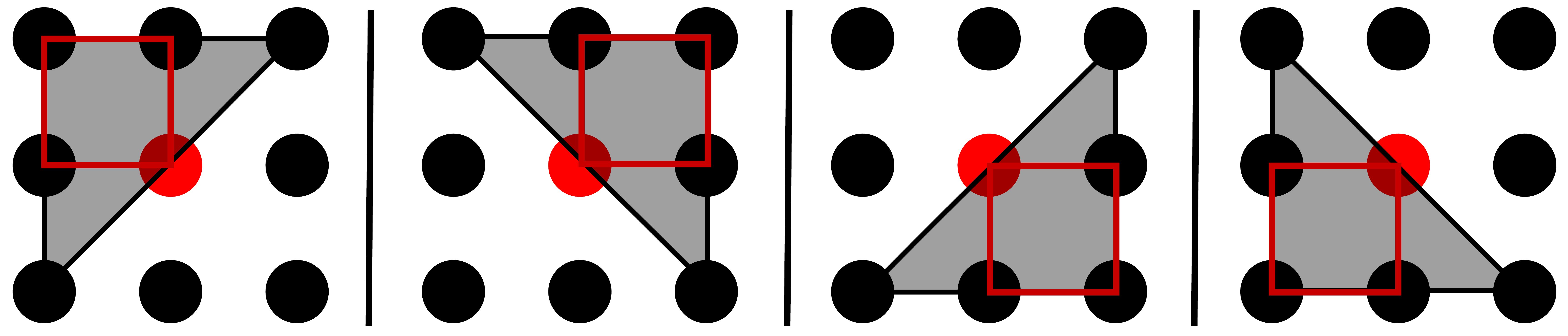}}
	\caption{Diagonal triangular half windows are ignored because they can be approximated by the corresponding quarter sub windows. The quarter support regions are smaller and also more computationally efficient.}
	\label{fig:diag4}
\end{figure}

In our filter, we ignore the four diagonal triangular half windows as shown in Fig.~\ref{fig:diag4}(a). There are two reasons. First, these triangular half windows can be approximated by the quarter sub windows as shown in Fig.~\ref{fig:diag4}(b) (they are also in top row of Fig.~\ref{fig:kernels8}). The second reason comes from the computational perspective. The cases in Fig.~\ref{fig:diag4}(a) are not separable, which needs more computation. In contrast, all the cases in Fig.~\ref{fig:kernels8} can be implemented by separable 1D kernels, which is more computationally efficient. Based on these two reasons, we ignore the diagonal triangular half windows.

There are several benefits of performing sub window regression. First, these sub window shapes do not change arbitrarily as scale adaptive (Fig.~\ref{fig:illus}(c)) or cross (Fig.~\ref{fig:illus}(d)). Therefore, we do not have to deal with the complex changing of support regions. Instead, we have only eight possible support regions (Fig.~\ref{fig:kernels8}). Second, the $\Phi$ for these eight cases are independent from image content. This makes our filter computational efficient. Third, this sub window regression is valid for generic $\Phi$. Thus these sub windows will work for many well-known $\Phi$ in this field. 

\subsection{Our Contribution}
The fundamental assumption in this paper is that every location $(x,y)$ in the image is at a corner or on an edge. This assumption is not well-known in the field and is not intuitive at first glance. When $(x,y)$ is at a flat region, it is also on the edge where left plane and right plane live on the same plane. Therefore, flat region is a special edge.

This assumption fundamentally changes the geometric role of location $(x,y)$. And only corners and edges need to be considered for each location in domain $\Omega$. Our assumption simplifies all possible configurations to only eight cases that can be easily analyzed, as shown in later sections.  

We perform the one-sided window regression with the constant weight function, leading to a one-sided box filter. The one-sided box filter can smooth the image but preserve edges. The presented filter has following properties
\begin{itemize}
	\item The presented one-sided box filter has $O(1)$ computational complexity with respect to window radius. 
	\item The presented one-sided box filter has linear computational complexity with respect to total number of pixels.
	\item The presented one-sided box filter can preserve corners and edges during the smoothing.
\end{itemize}

\section{one-sided filter}
In this section, we first show why we should perform sub window regression from function analysis point of view. Then, we show that regression on several sub-windows is a generic idea and is valid for generic $\Phi$. We will discuss the one-sided box filter and its properties in next section.
\subsection{One-Sided Support Regions}
\begin{figure}
	\centering
	\includegraphics[width=0.7\linewidth]{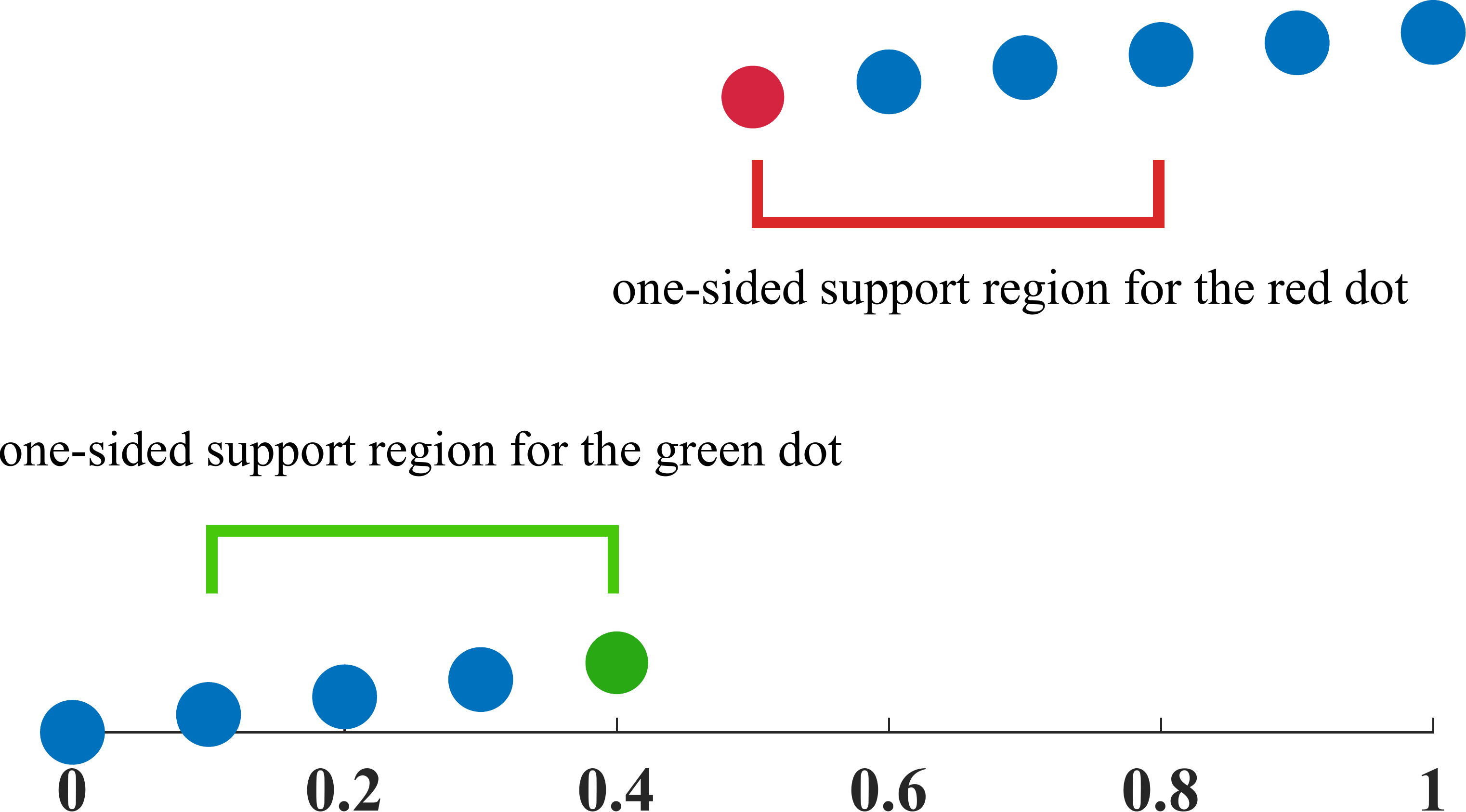}
	\caption{In the one dimensional case, at the discontinuous location (indicated by green and red circles), their support regions should be one-sided.}
	\label{fig:oneD}
\end{figure}

Our one-sided support idea comes from the one-sided limits in calculus. As shown in Fig.~\ref{fig:oneD}, one-sided limits indicate that the support region should come from one side when there is a discontinuity in the function. Such discontinuity location indicates the edge in the image. This fact inspires us to restrict the support region at the one side of the given location. 

As shown in Fig.~\ref{fig:oneD}, although the green and red dots are spatially close to each other in the imaging domain (0.4 and 0.5), they belong to different classes from function value (image intensity) point of view. In the 1D case, one location can only belong to left or right interval, as indicated by the colored lines in Fig.~\ref{fig:oneD}.

Such observation can be theoretically explained from Taylor expansion point of view. Let $U(x)$ denote the one dimensional signal, and $x_g, x_r$ be the green dot and red dot location, respectively. Then, for a small positive $\epsilon>0$, from Taylor expansion, we have
\begin{eqnarray}
U(x_g)\approx U(x_g-\epsilon)+U'(x_g-\epsilon)\epsilon\,\\
U(x_r)\approx U(x_r+\epsilon)-U'(x_r+\epsilon)\epsilon\,.
\end{eqnarray} Due to the discontinuity $U(x_g)\neq U(x_r)$, such approximation can not cross the discontinuity location. Therefore, the support region can only be one-sided.

In the 2D case, the one-sided region is more complex. In 2D, the discontinuity of $U(x,y)$ is the edge jump, which has the following normal direction
\begin{equation}
\label{eq:normal}
	\vec{n}=\frac{\nabla U}{\|\nabla U\|_2}\,.
\end{equation} Different from the 1D case that only has two directions (left and right, see Fig.~\ref{fig:oneD}), the 2D normal has infinite number of possible directions. Despite this issue, another problem is that $\|\nabla U\|$ might be zero in the denominator, leading to the numerical instability in above equation. In fact, $\|\nabla U\|=0$ has the highest probability for natural images~\cite{gong:gdp}. Therefore, directly computing the normal directions by this equation is neither accurate nor numerical stable. 

\begin{figure}[!htb]
	\centering
	\includegraphics[width=0.8\linewidth]{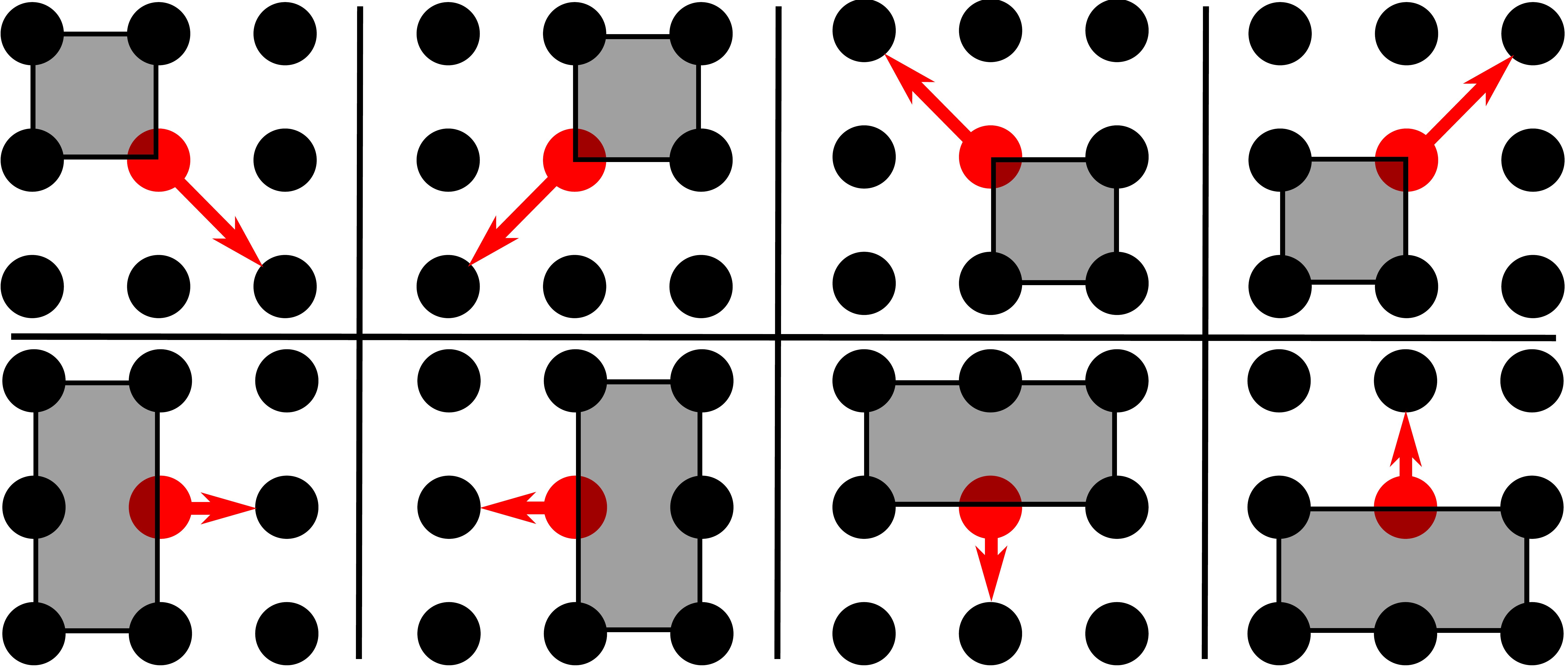}
	\caption{Eight directions for the one-sided sub window. The red arrow from the center location indicates the direction. Each direction is corresponding to the shaded sub window.}
	\label{fig:normal}
\end{figure}

Luckily, we can use eight discrete directions $\vec{n}_i$ to approximate $\vec{n}$ in a local window~\cite{GONG2019329}. As shown in Fig.~\ref{fig:normal}, the red arrow indicates the normal direction and the shaded sub window region indicates the one-sided support region for the corresponding normal direction. Although these eight discrete directions do not cover all possible normal directions, they are good approximations to the real normal direction $\vec{n}$~\cite{GONG2019329}. The maximum angle error is $\frac{\pi}{8}$, which does not cause obvious visual differences. Moreover, such enumerable support regions avoid the numerical issue $\|\nabla U\|=0$. Unlike Eq.~\ref{eq:normal} only valid for large gradients,  these enumerable normal directions are valid for the whole imaging domain $\Omega$.

We define the one-sided support regions (window) for location $(x,y)$ as the eight sub-windows defined by Eq.~(6) to Eq.~(13). The first four sub windows assume that the central location $(x,y)$ is at a corner. The rest sub windows assume that the central location is on an edge. Such assumption is also valid when the central location is in a flat region.
\begin{eqnarray}
{\cal N}_1(x,y)=\{(u,v)|x-r\le u\le x\,,y\le v\le y+r\}\\
{\cal N}_2(x,y)=\{(u,v)|x\le u\le x+r\,,y\le v\le y+r\}\\
{\cal N}_3(x,y)=\{(u,v)|x\le u\le x+r\,,y-r\le v\le y\}\\
{\cal N}_4(x,y)=\{(u,v)|x-r\le u\le x\,,y-r\le v\le y\}\\
{\cal N}_5(x,y)=\{(u,v)|x-r\le u\le x\,,|y-v|\le r\}\\
{\cal N}_6(x,y)=\{(u,v)|x\le u\le x+r\,,|y-v|\le r\}\\
{\cal N}_7(x,y)=\{(u,v)||u-x|\le r,y\le v\leq y+ r\}\\
{\cal N}_8(x,y)=\{(u,v)||u-x|\le r,y-r\le v\leq y\}\,.
\end{eqnarray}

We use the closest distance to choose one sub window. When $(x,y)$ is in a flat region, choosing which sub window is not important. But when $(x,y)$ is the edge or corner location, we must choose the correct sub window to perform $\Phi$. In order to automatically make this decision, we run $\Phi$ on all eight sub windows and get eight estimations. Then we select the closest one as the result. This method is named as One-Sided Filter. And it is summarized in Algorithm~\ref{algo:HW}.

\begin{algorithm}[!htb]
	\caption{One-Sided Filter}
	\label{algo:HW}
	\begin{algorithmic}
		\Require $I(x,y)$, window radius $r$, $\Phi$, IterationNum
		\State $t=0$, $U^0(x,y)=I(x,y)$
		\While{$t \leq$ IterationNum}
		\For{$(x,y)\in\Omega$}
		\State $d_i=\Phi(U^{t}({\cal N}_i(x,y))) - U^{t}(x,y),\,~i\in\{1,\cdots,8\}$ 
		\State $m=\arg\min\limits_{i\in\{1,\cdots,8\}}\{|d_i|\}$
		\State $U^{t+1}(x,y) = U^{t}(x,y) + d_m(x,y)$
		\EndFor
		\State $t=t+1$
		\EndWhile
		\Ensure $U(x,y)$
	\end{algorithmic}
\end{algorithm}

\begin{figure}[!htb]
	\centering
	\includegraphics[width=0.23\linewidth]{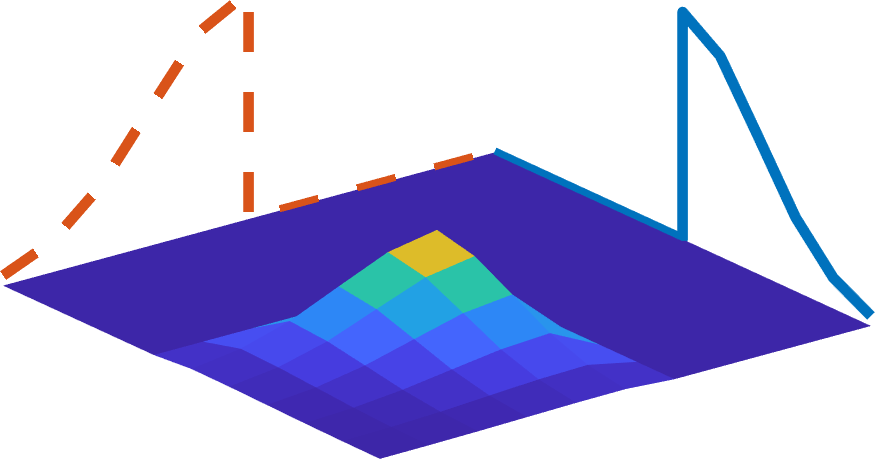}
	\includegraphics[width=0.23\linewidth]{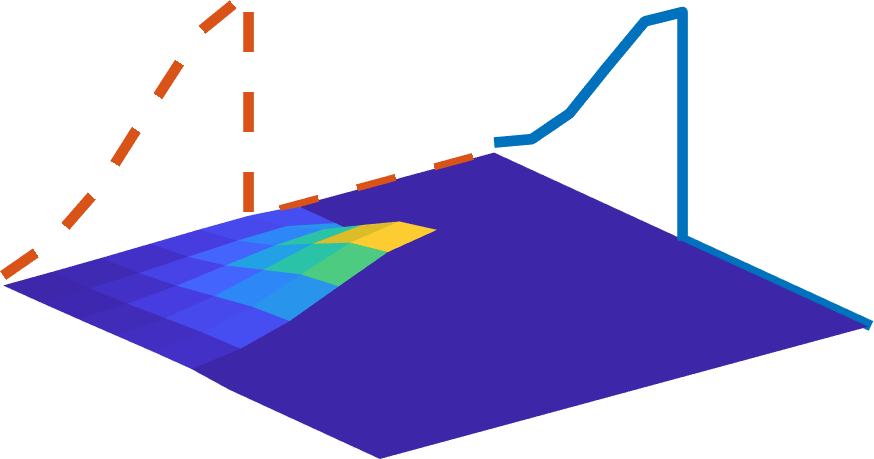}
	\includegraphics[width=0.23\linewidth]{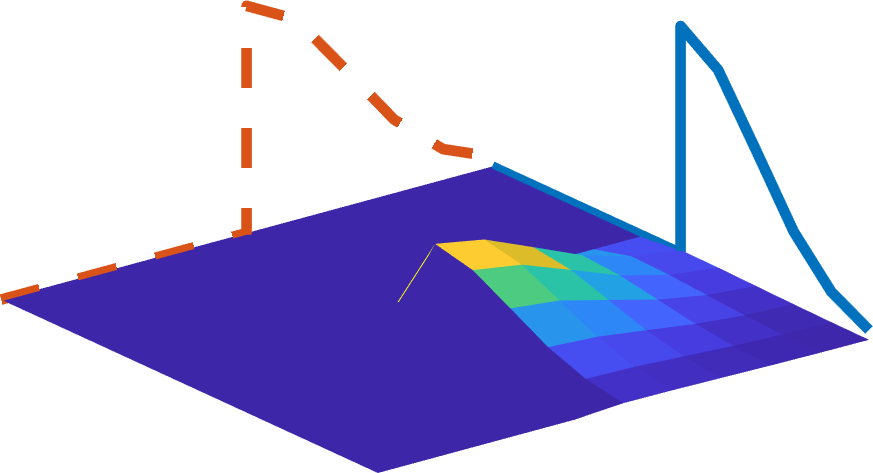}
	\includegraphics[width=0.23\linewidth]{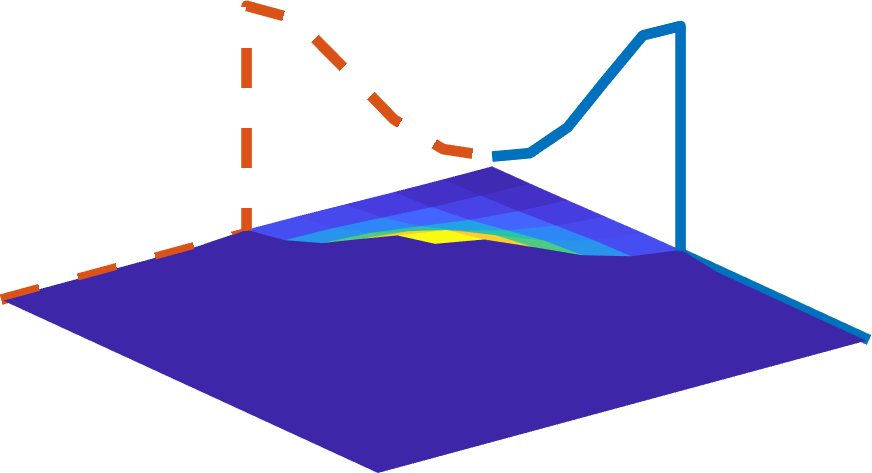}
	\includegraphics[width=0.23\linewidth]{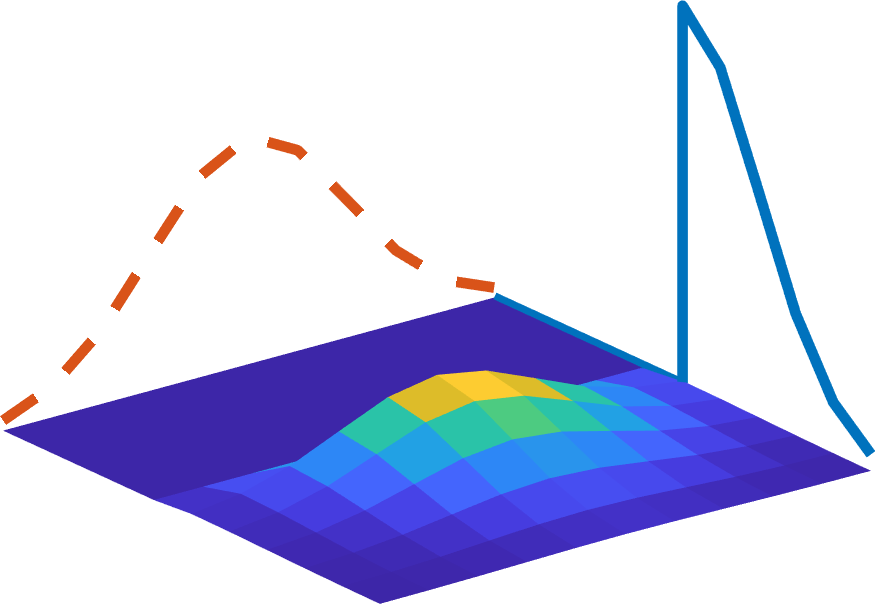}
	\includegraphics[width=0.23\linewidth]{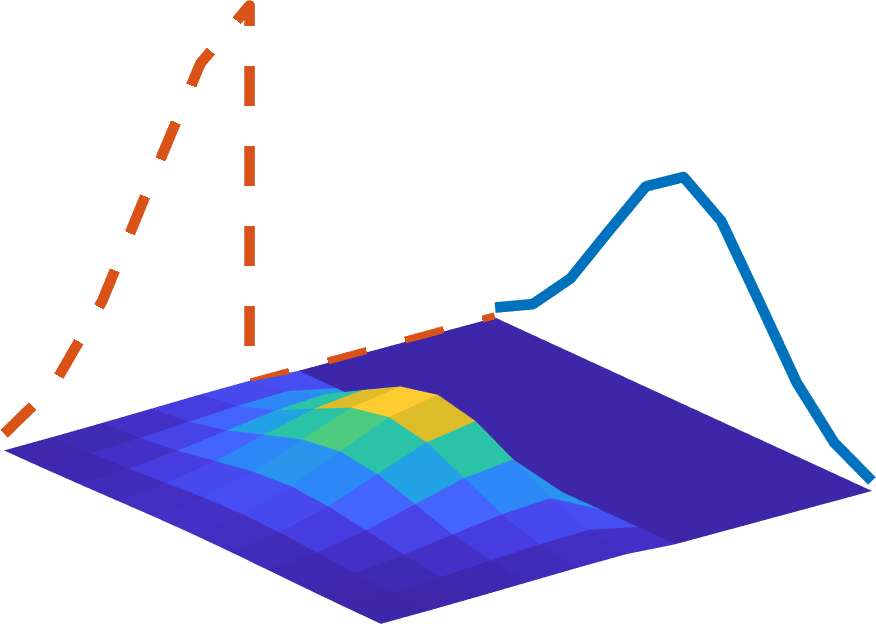}
	\includegraphics[width=0.23\linewidth]{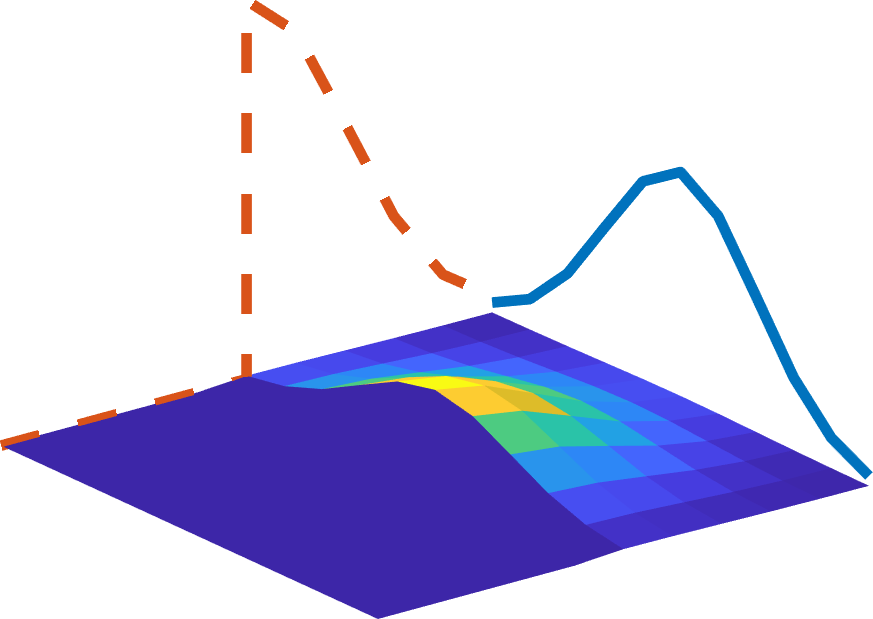}
	\includegraphics[width=0.23\linewidth]{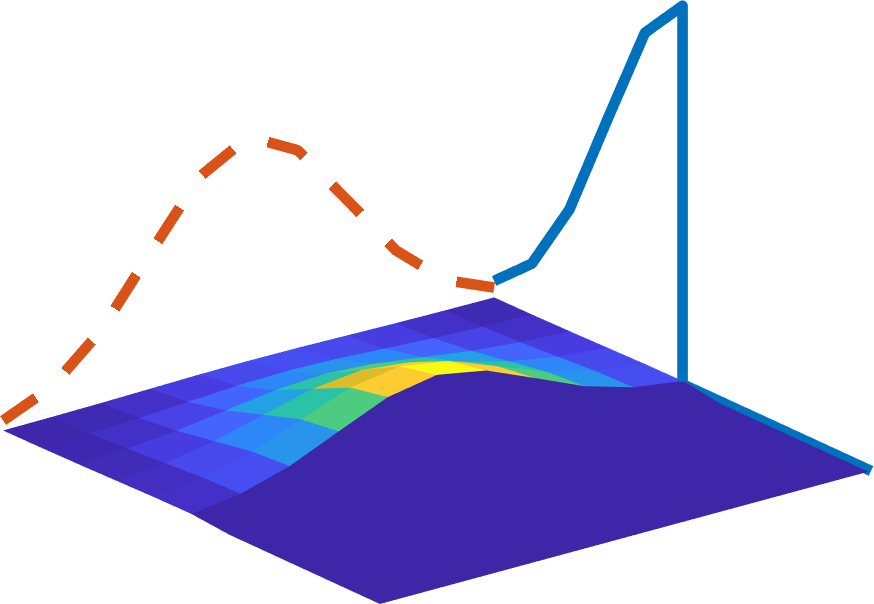}
	
	\subfigure{
		\centering
		\includegraphics[width=0.22\linewidth]{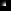}
		\includegraphics[width=0.22\linewidth]{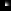}
		\includegraphics[width=0.22\linewidth]{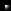}
		\includegraphics[width=0.22\linewidth]{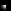}}
	\subfigure{
		\centering
		\includegraphics[width=0.22\linewidth]{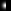}
		\includegraphics[width=0.22\linewidth]{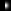}
		\includegraphics[width=0.22\linewidth]{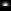}
		\includegraphics[width=0.22\linewidth]{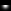}}
	\caption{Eight 2D one-sided kernels on ${\cal N}$ with Gaussian weight function($r=5$). Top two rows: the dash red line and solid blue line show the separated 1D kernels. Bottom two rows: corresponding one-sided Gaussian convolution kernels.}
	\label{fig:kernels}
\end{figure}

One-sided filter can be considered as anisotropic diffusion and the diffusion only happens in the sub window. Therefore, it can preserve edges and corners for arbitrary weight function $\Phi$. And such preserving is theoretically guaranteed, without manually tuning any parameter.
\subsection{Separable Kernels}
In this section, we show how to construct the one-sided kernels from traditional symmetric separable kernels. As mentioned, the local smoothing methods usually take following form
\begin{equation}
U(x,y)=\int\limits_{u=-r}^{r}\int\limits_{v=-r}^{r}k(u,v)I(x-u,y-v)\mathrm{d}u\mathrm{d}v\,.
\end{equation} When $k$ is separable, it can be expressed as
\begin{equation}
k(u,v)=k_x(u)k_y(v)\,,
\end{equation} where $k_x$ and $k_y$ are separated kernels in $x$ and $y$ direction, respectively.

Now, we show how to construct the one-sided kernels in our method. Once we have $k_x(u)$, we can construct the normalized one-sided kernels for $x$ direction
\begin{equation}
\label{eq:oneside}
k^-_x(u)=\frac{\tilde{k}_x^-(u)}{\int\tilde{k}_x^-(u)\mathrm{d}u}~~\mathrm{and}~~k^+_x(u)=\frac{\tilde{k}_x^+(u)}{\int\tilde{k}_x^+(u)\mathrm{d}u}\,,
\end{equation} where
\begin{eqnarray}
\tilde{k}_x^-=
\begin{cases}
k_x(u), &u\le 0 \cr 0, &u>0
\end{cases}~~\mathrm{and}~~
\tilde{k}_x^+=
\begin{cases}
0, &u\le 0 \cr k_x(u), &u>0
\end{cases}\,.
\end{eqnarray} Similarly, we can have normalized one-sided kernels $k^-_y(v)$ and $k^+_y(v)$ for $y$ direction. The four 1D kernels $\{k_x^-,k_x^+, k_y^-, k_y^+\}$ span the 2D kernels in the first row of Fig.~\ref{fig:kernels8}. Combining these half kernels with $k_x$ or $k_y$, we have the half window 2D kernels in the second row of Fig.~\ref{fig:kernels8}. 

One example of such kernels are shown in Fig.~\ref{fig:kernels}, where the classical 2D Gaussian weight function is adopted. The first two rows show the normalized separated 1D kernels and the bottom two rows show the corresponding spanned 2D convolution kernels. The anisotropic diffusion property of these kernels becomes clear.

\section{one-sided box filter}
In this section, we show the one-sided filter with the constant function, leading to the one-sided box filter. This filter can be interpreted as a special case of one-sided Gaussian filter ($\sigma=+\infty$). Such relationship comes from the relationship between the classical box filter and Gaussian filter. 

We focus on one-sided box filter (instead of one-sided Gaussian filter) for several reasons. First, this one-sided box filter can preserve edges and corner. Second, this one-sided box filter inherits the $O(1)$ computational complexity of original box filter with respect to the window size and also the linear $O(N)$ computational complexity with respect to the total number of pixels. Third, this one-sided box filter has a fast implementation and can achieve high performance on practical applications for high resolution images.  

It is well-known that box filter can be implemented by the integral image technique. This makes the box filter has $O(1)$ computational complexity with respect to the window radius. Such complexity leads to the high performance of box filter and further popularizes its application on practical tasks. 

In this section, we will show that one-sided box filter inherits such properties. It has the same computational complexity as the original box filter. But it can preserve edges and corners much better. Therefore, we believe the presented filter will impact on a large range of applications.
\subsection{One-sided Kernels for Box Filter}
Combining the sub window regions and the constant weight function, we can define the one-sided kernels with the constant weight function. More specifically, we have following eight kernels (also reported in our preliminary work~\cite{Gong2018})
\begin{eqnarray}
k_1=&
\begin{cases}
(r+1)^{-2}, &-r\leq u\leq 0,\,0\leq v\leq r \cr 0, & else
\end{cases}\\
k_2=&
\begin{cases}
(r+1)^{-2}, &0\leq u\leq r,\,0\leq v\leq r \cr 0, & else
\end{cases}\\
k_3=&
\begin{cases}
(r+1)^{-2}, &0\leq u\leq r,\,-r\leq v\leq 0 \cr 0, & else
\end{cases}\\
k_4=&
\begin{cases}
(r+1)^{-2}, &-r\leq u\leq 0,\,-r\leq v\leq 0 \cr 0, & else
\end{cases}\\
k_5=&
\begin{cases}
(r+1)^{-1}(2r+1)^{-1}, &-r\leq u\leq 0, \cr 0, & else
\end{cases}
\end{eqnarray}
\begin{eqnarray}
k_6=&
\begin{cases}
(r+1)^{-1}(2r+1)^{-1}, &0\leq u\leq r, \cr 0, & else
\end{cases}\\
k_7=&
\begin{cases}
(r+1)^{-1}(2r+1)^{-1}, &0\leq v\leq r, \cr 0, & else
\end{cases}\\
k_8=&
\begin{cases}
(r+1)^{-1}(2r+1)^{-1}, &-r\leq v\leq 0 \cr 0, & else
\end{cases}\,.
\end{eqnarray}
Each kernel $k_i$ corresponds to a one-sided sub window ${\cal N}_i$ in Fig.~\ref{fig:normal}. Since the original box filter kernel is separable, each kernel in the one-sided box filter is also separable. More specifically, we take the constant weight function into the general one-sided filter (Eq.~\ref{eq:oneside})
\begin{equation}
	k_{1x}=\frac{1}{r+1},-r\leq u\leq 0\,,k_{1y}=\frac{1}{r+1},0\leq v\leq r\,.
\end{equation} Other separable kernels for $k_{2,3,4,5,6,7,8}$ can be found in similar way. These separable kernels lead to a faster approximation implementation (Section~\ref{sec:fast}) and an extension for higher dimensional data (Section~\ref{sec:high}).

The algorithm is summarized in Algorithm.~\ref{algo:HWB}. The implementation of this one-sided box filter is straightforward. And our MATLAB implementation of this algorithm with separable kernels can be found in the appendix. 
\begin{algorithm}[!htb]
	\caption{One-Sided Box Filter}
	\label{algo:HWB}
	\begin{algorithmic}
		\Require $I(x,y)$, window radius $r$, IterationNum
		\State $t=0$, $U^0(x,y)=I(x,y)$
		\While{$t \leq$ IterationNum}
		
		\State $d_i=k_i\otimes U^{t}-U^{t},\,~i\in\{1,\cdots,8\}$ 
		\State $m=\arg\min\limits_{i\in\{1,\cdots,8\}}\{|d_i|\}$
		\State $U^{t+1} = U^{t} + d_m$
		\State $t=t+1$
		\EndWhile
	\end{algorithmic}
\end{algorithm}

\begin{figure}[!htb]
	\centering
	{\centering ~~~~~~Iteration=2~~~~~~~~~ Iteration=10~~~~~~~~~ Iteration=30}
	\rotatebox{90}{\centering\hspace{1cm} $r$=10\hspace{2cm} $r$=5\hspace{2cm} $r$=2}~~\includegraphics[width=0.9\linewidth]{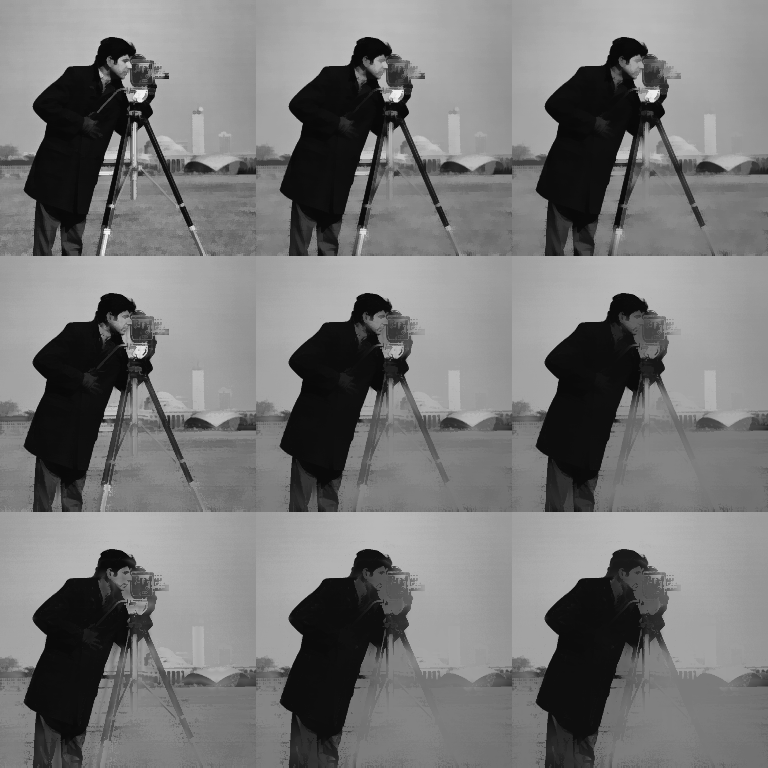}
	\caption{One-sided box filter with different window radius and iterations. The window radius and iteration number play different roles in this algorithm.}
	\label{fig:parameters}
\end{figure}

\begin{figure}[!tbh]
	\centering
	\includegraphics[width=0.7\linewidth]{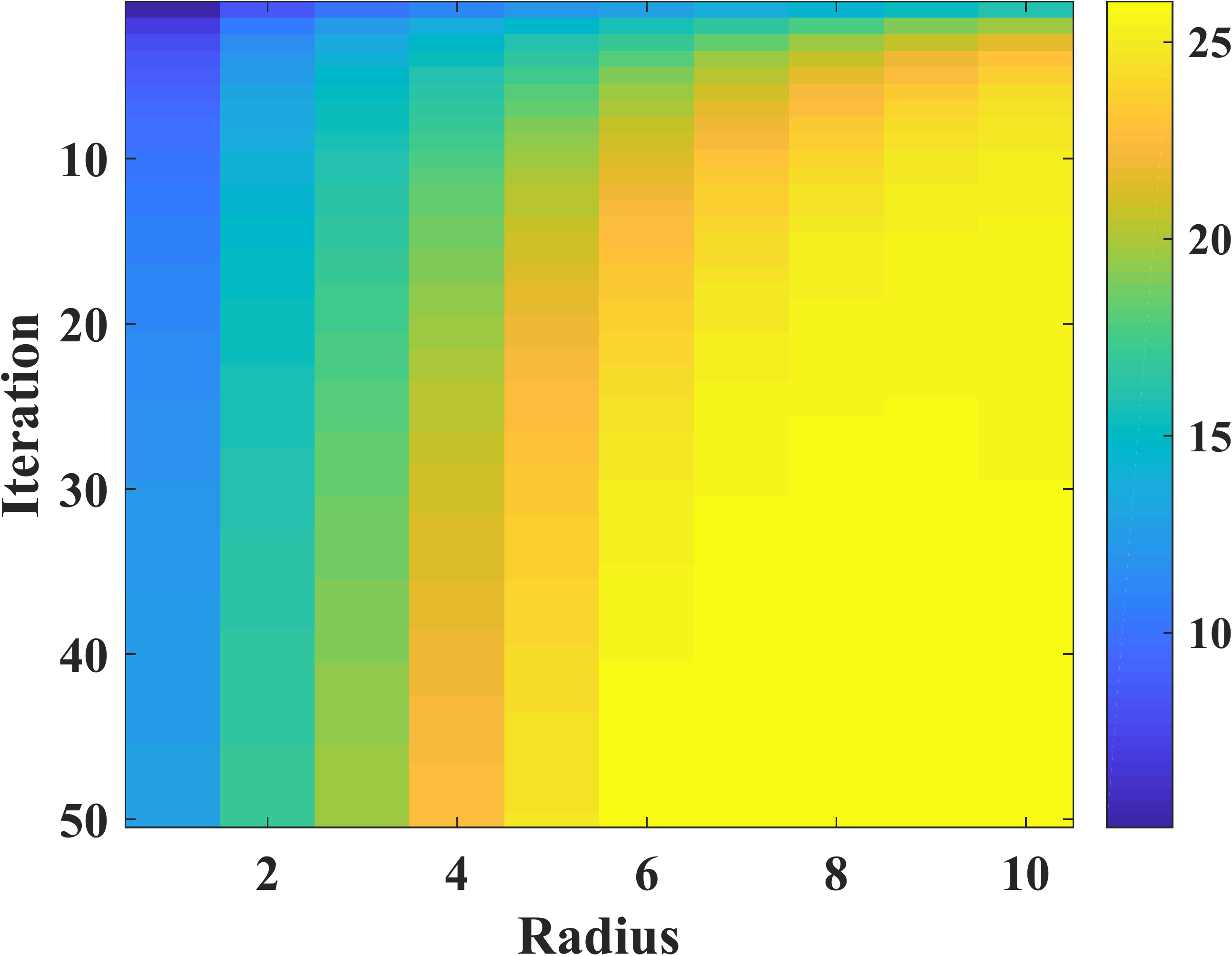}
	\caption{The RMSE between the original and smoothed. The smaller window radius needs the larger iteration number to converge.}
	\label{fig:parameters2}
\end{figure}

\subsection{Parameter Analysis}
In this algorithm, the window radius and iteration number play different roles. This property is different from the classical box filter where the window raidus and iteration number play similar role (increasing window radius is equivalent with increasing iteration number). To show this property, we show one example in Fig.~\ref{fig:parameters} and the corresponding Root Mean Squared Error (RMSE) is shown in Fig.~\ref{fig:parameters2}.

From Fig.~\ref{fig:parameters}, we can tell that the window radius and iteration number play different roles in this algorithm. Increasing window radius is NOT equivalent with increasing the iteration number. This property is different from the classical box filter.

From Fig.~\ref{fig:parameters2}, we can also tell that the larger radius needs the smaller iteration number to converge. This can be explained from diffusion point of view. The diffusion with larger window radius leads to long range information propagation. Therefore, it reaches the steady state in a few iterations. In contrast, the algorithm with the smaller window radius needs the larger iteration numbers to converge.

\subsection{Frequency Response Analysis}
Computing $d_i$ in Algorithm~\ref{algo:HWB} is important for this filter. Such computation can be carried out by a simple convolution operator with kernel $k_i-\delta$, where $\delta$ is a function
\begin{equation}
	\delta(\vec{x})=\begin{cases}
	1, &\vec{x}=0 \cr 0, & else
	\end{cases}\,.
\end{equation}

\begin{figure}
	\centering
	\includegraphics[width=0.2\linewidth]{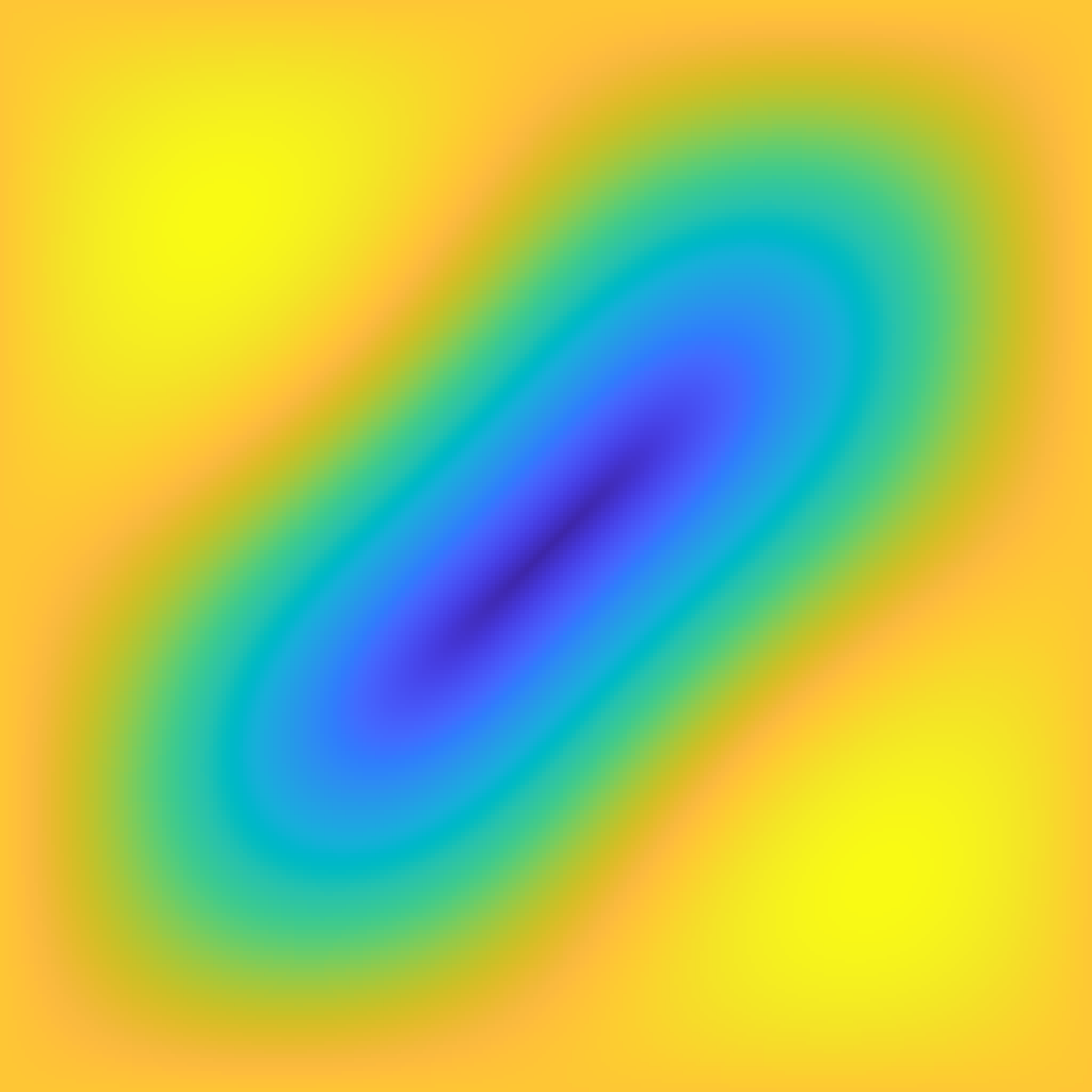}
	\includegraphics[width=0.2\linewidth]{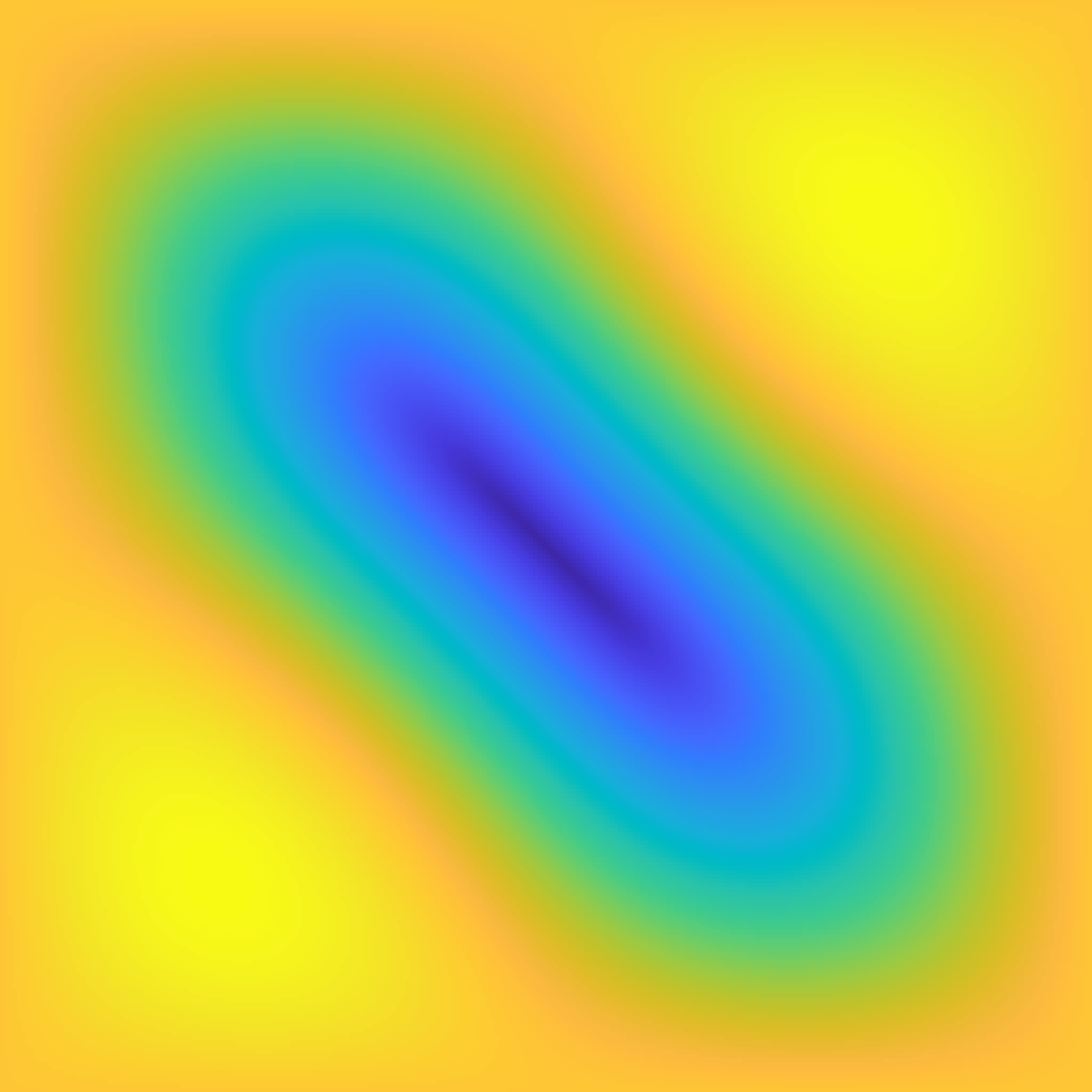}
	\includegraphics[width=0.2\linewidth]{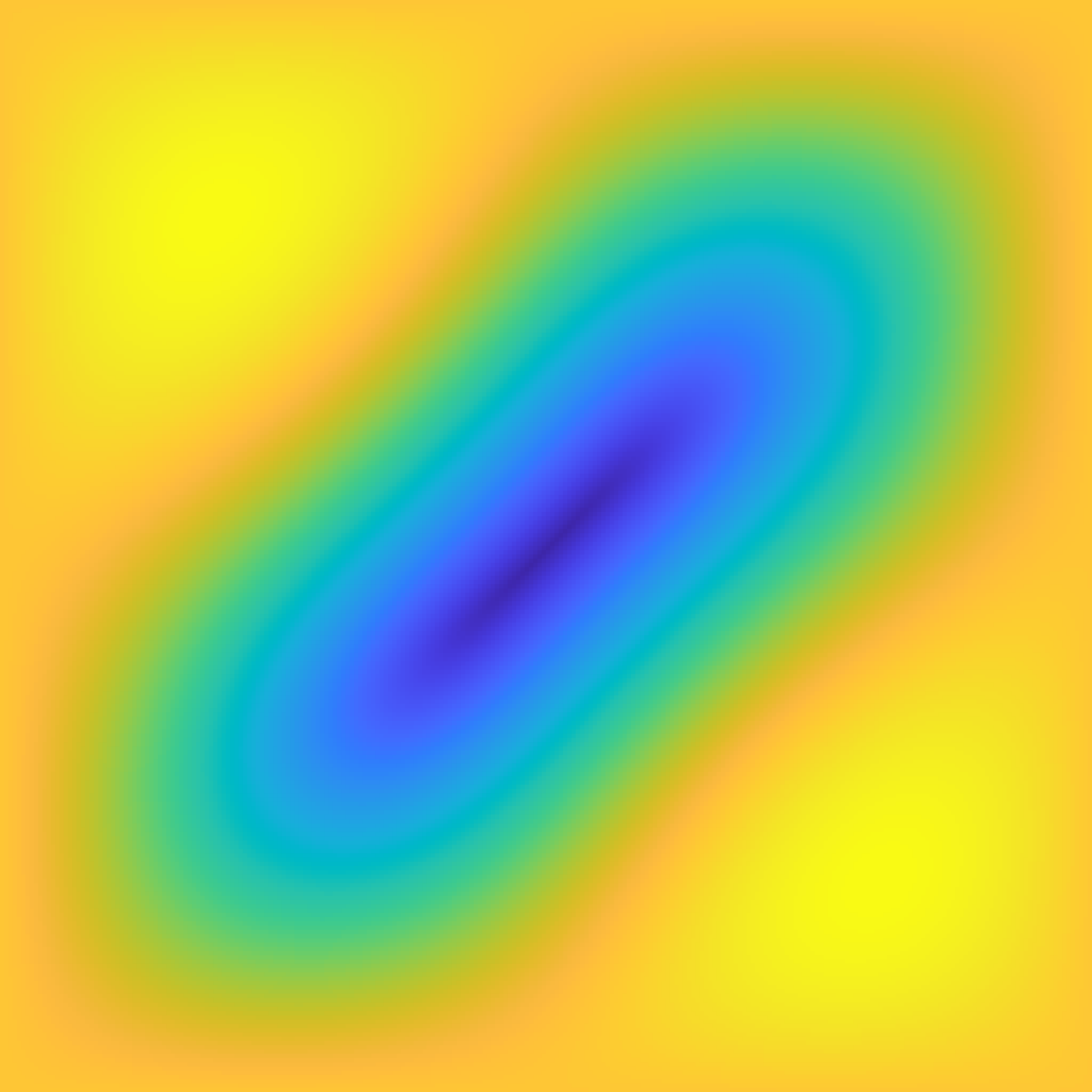}
	\includegraphics[width=0.2\linewidth]{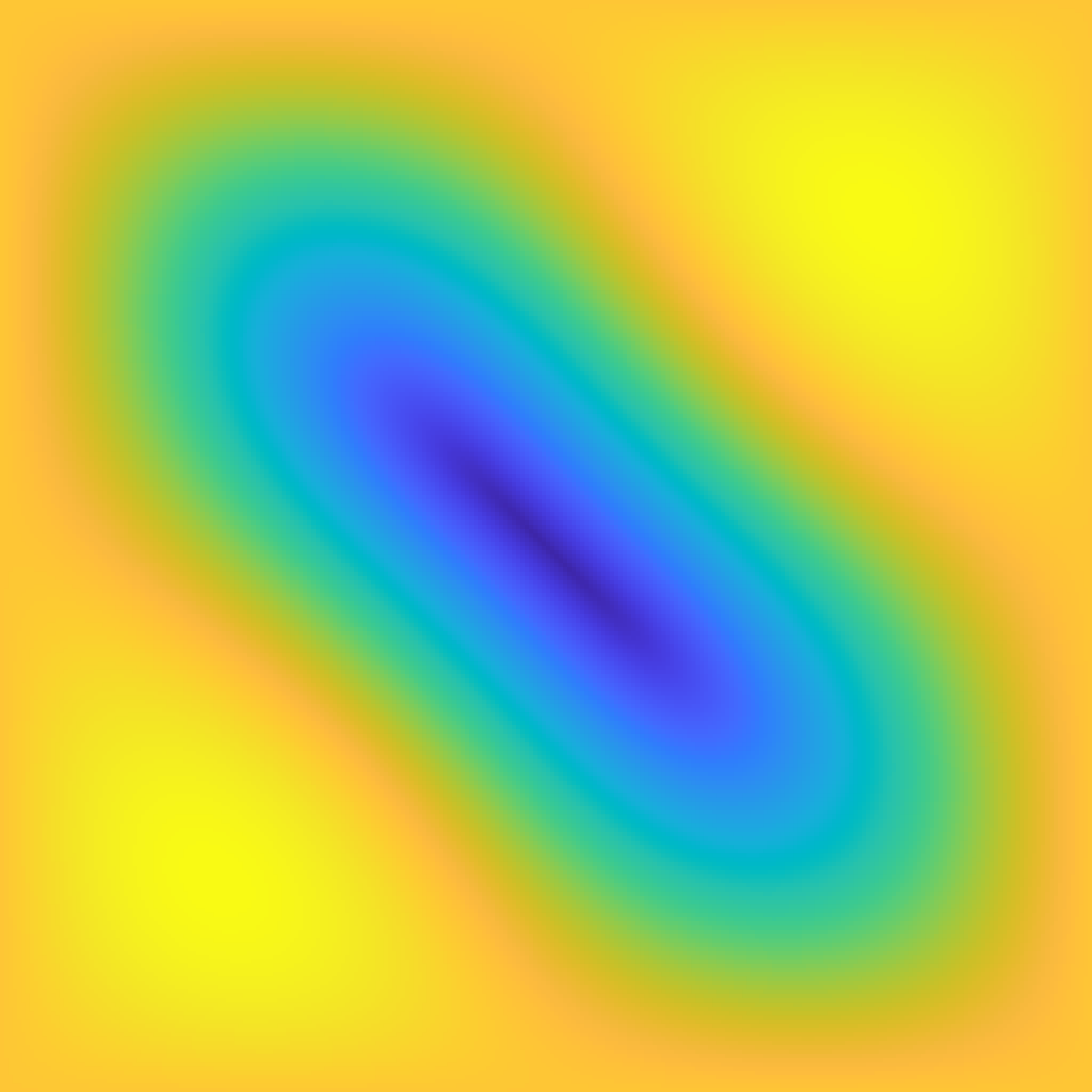}\vspace{1.5mm}
	\includegraphics[width=0.2\linewidth]{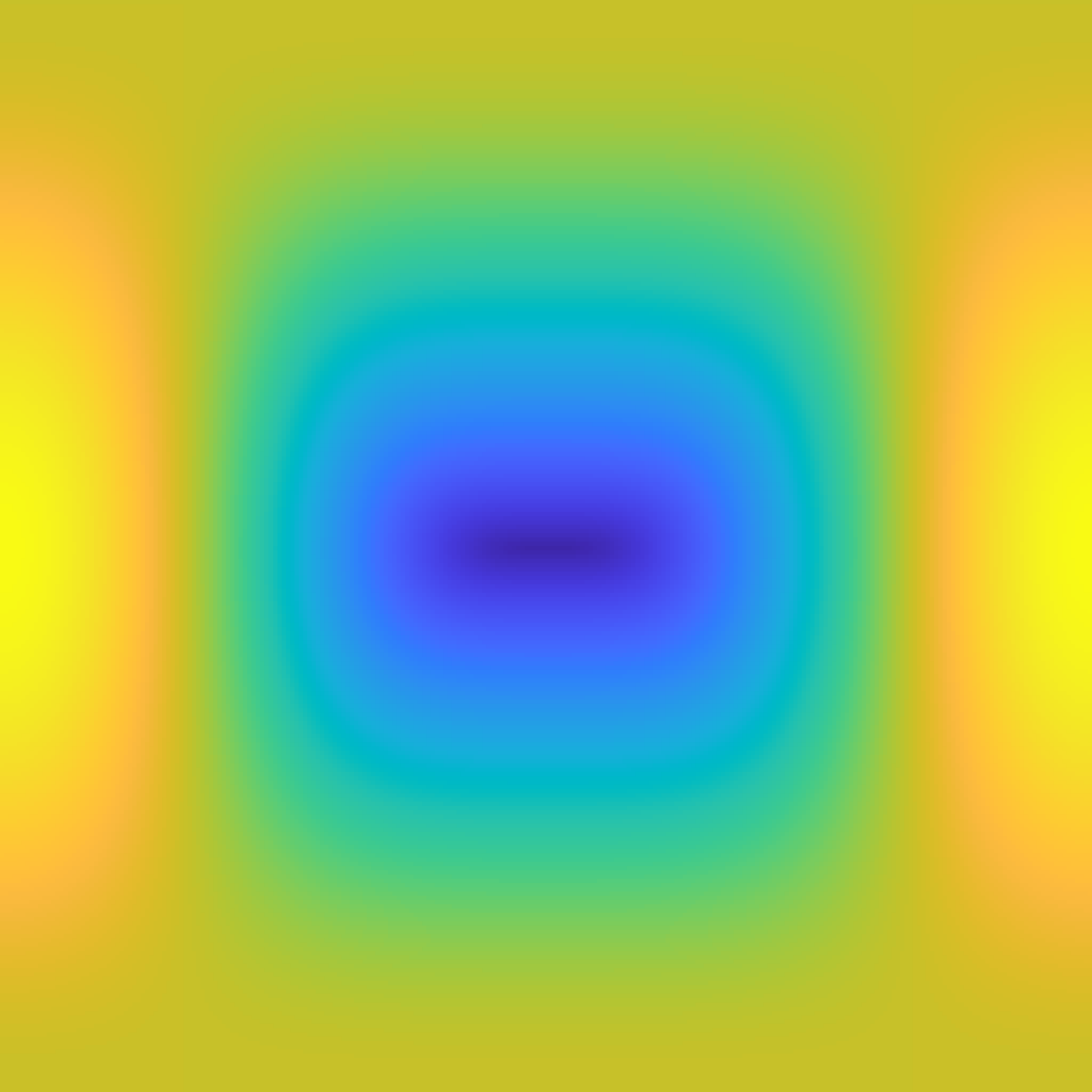}
	\includegraphics[width=0.2\linewidth]{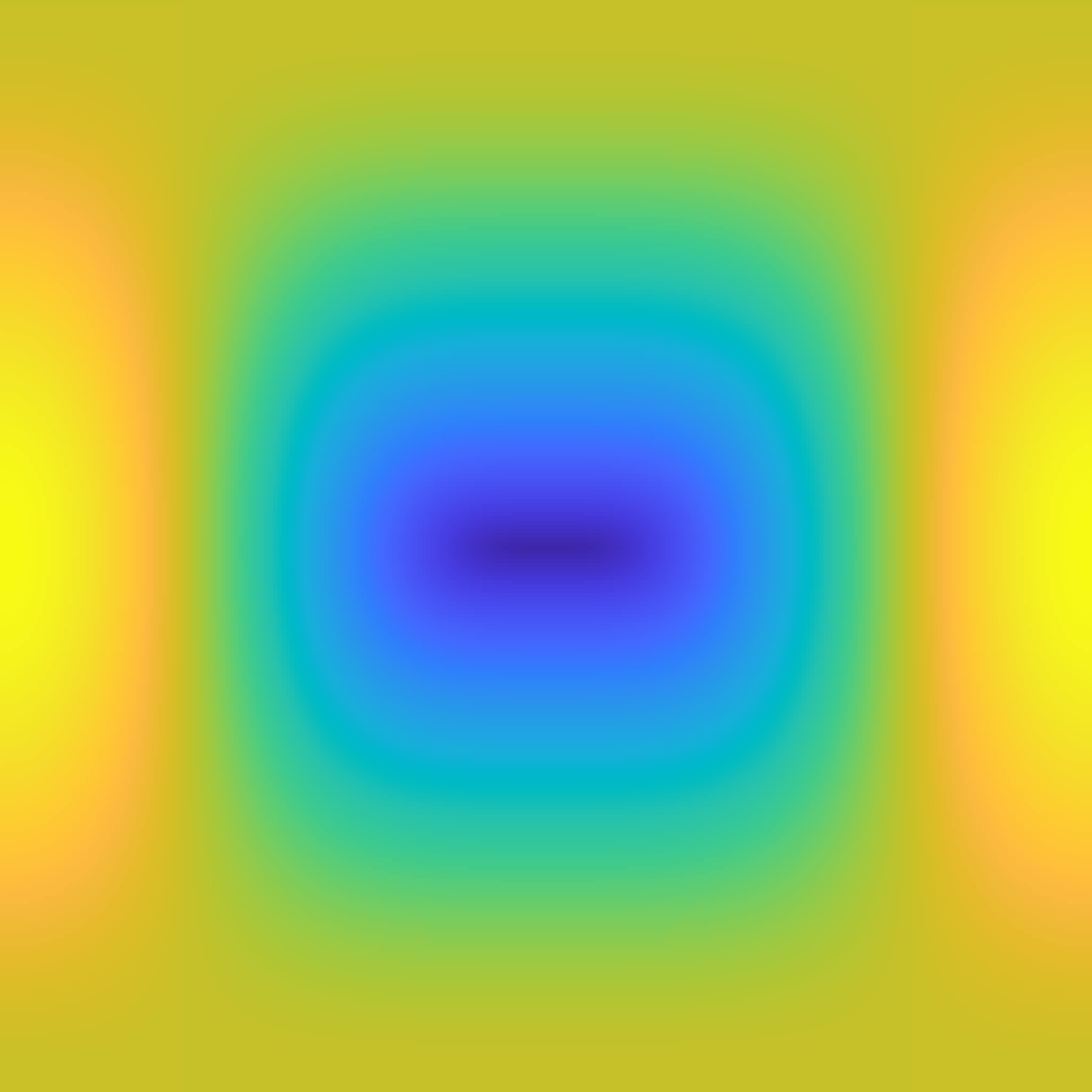}
	\includegraphics[width=0.2\linewidth]{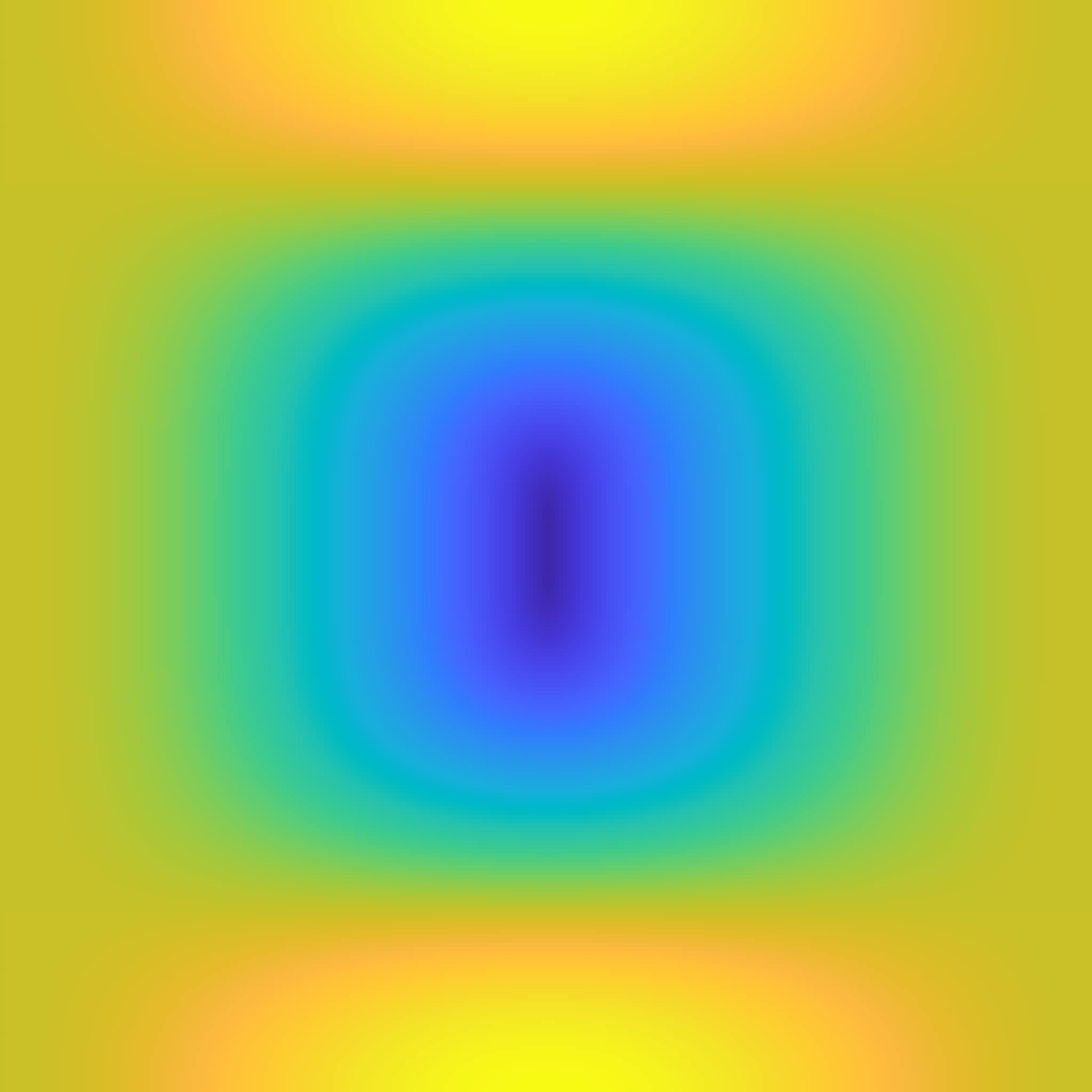}
	\includegraphics[width=0.2\linewidth]{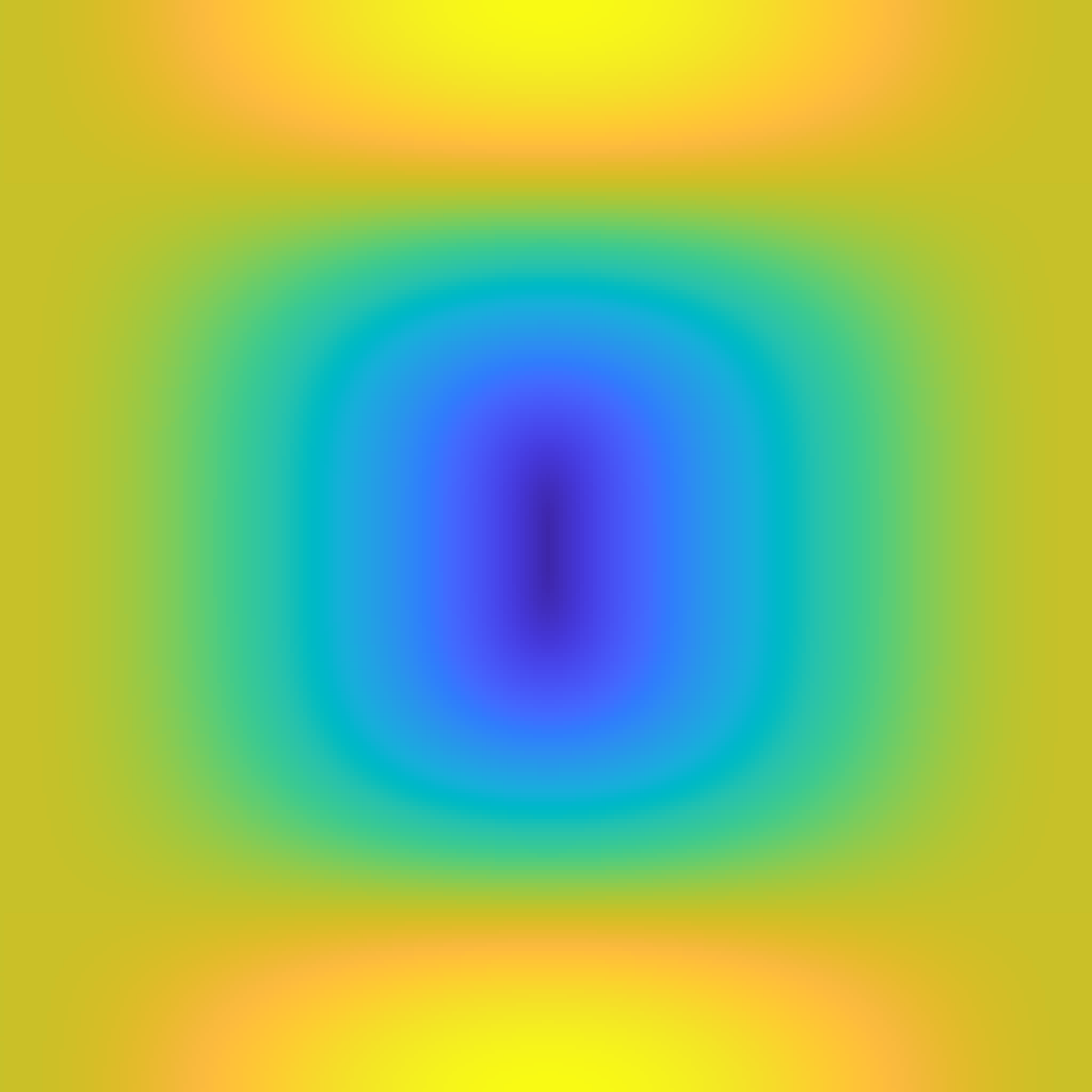}
	\caption{Fourier spectrum of $k_i-\delta$ with window radius $r=1$. Yellow color indicates high value while blue color indicates low value.}
		\label{fig:fft}
\end{figure}

The Fourier spectrum (magnitude) for $k_i-\delta$ with $r=1$ is shown in Fig.~\ref{fig:fft}. Such spectrum confirms that $d_i$ can be interpreted as anisotropic diffusion process.

Meanwhile, the line $m=\arg\min\limits_{i\in\{1,\cdots,8\}}\{|d_i|\}$ in Algorithm~\ref{algo:HWB} provides the nonlinearity for this filter. Therefore, algorithm~\ref{algo:HWB} can be interpreted as a nonlinear anisotropic filter.
 
\subsection{$O(1)$ and $O(N)$ computational complexity}
The computational complexity of one-sided box filter is independent from the window radius $r$. Therefore, its computational complexity is $O(1)$ with respect to the window radius.

Further more, the computational complexity of one-sided box filter is the same as the computational complexity of the original box filter. Thus, it has linear computational complexity $O(N)$ with respect to the total number of samples $N$.

As mentioned, we focus on one-sided box filter because it is computationally efficient. Its computational complexity is the same as the original box filter. The running time of the Algorithm~\ref{algo:HWB}'s naive implementation is about eight times of the counterpart of the original box filter. Such computational complexity property is important for high resolution images, video processing and real time applications.

In later section, we will show a fast approximation of one-sided box filter. Although the fast approximation is not exactly the one-sided box filter, it leads to similar results as the one-sided box filter. But the running time of the fast approximation filter is about $60\%$ of the one-sided box filter running time.  

\subsection{Edge and Corner Preserving}
\begin{figure}
	\centering
	\subfigure[original]{\includegraphics[width=0.3\linewidth]{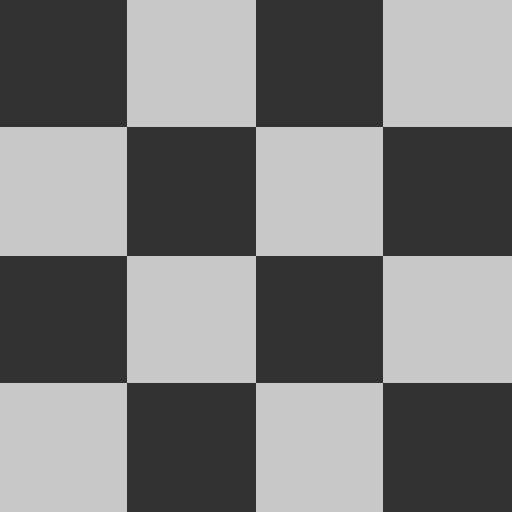}}
	\subfigure[noisy]{\includegraphics[width=0.3\linewidth]{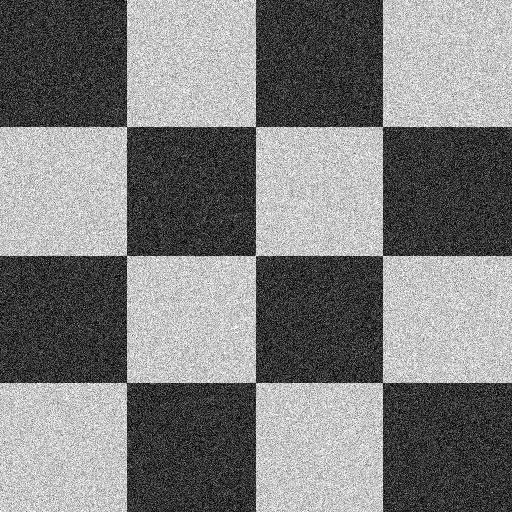}}
	\subfigure[Gaussian filter]{\includegraphics[width=0.3\linewidth]{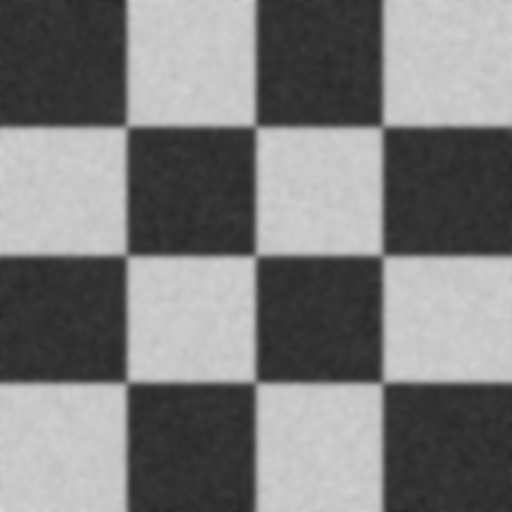}}
	\subfigure[Guided filter]{\includegraphics[width=0.3\linewidth]{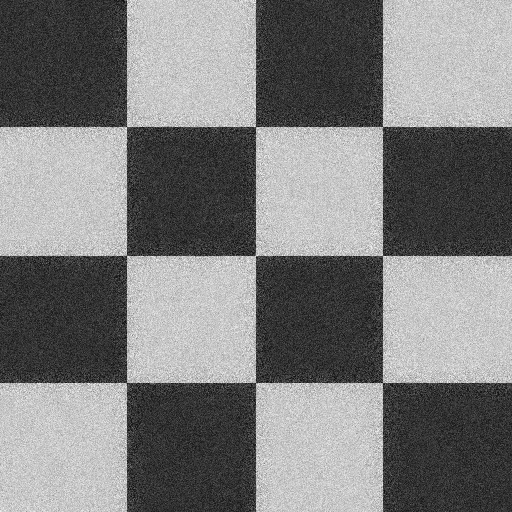}}
	\subfigure[one-sided Gaussian]{\includegraphics[width=0.3\linewidth]{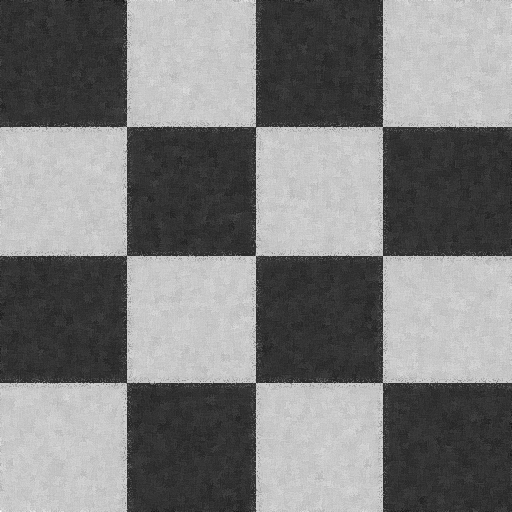}}
	\subfigure[one-sided box filter]{\includegraphics[width=0.3\linewidth]{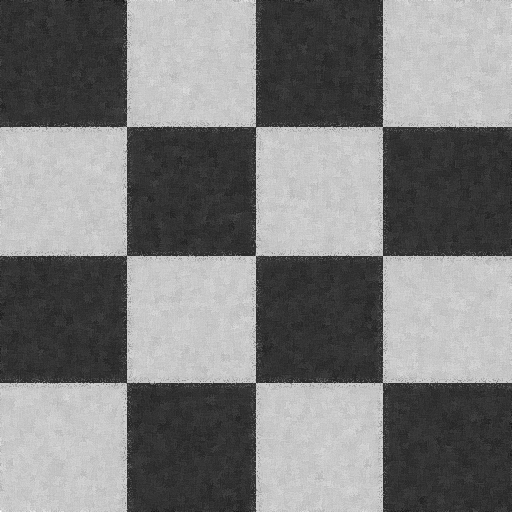}}
	
	\subfigure[PSNR for different radius]{\includegraphics[width=0.6\linewidth]{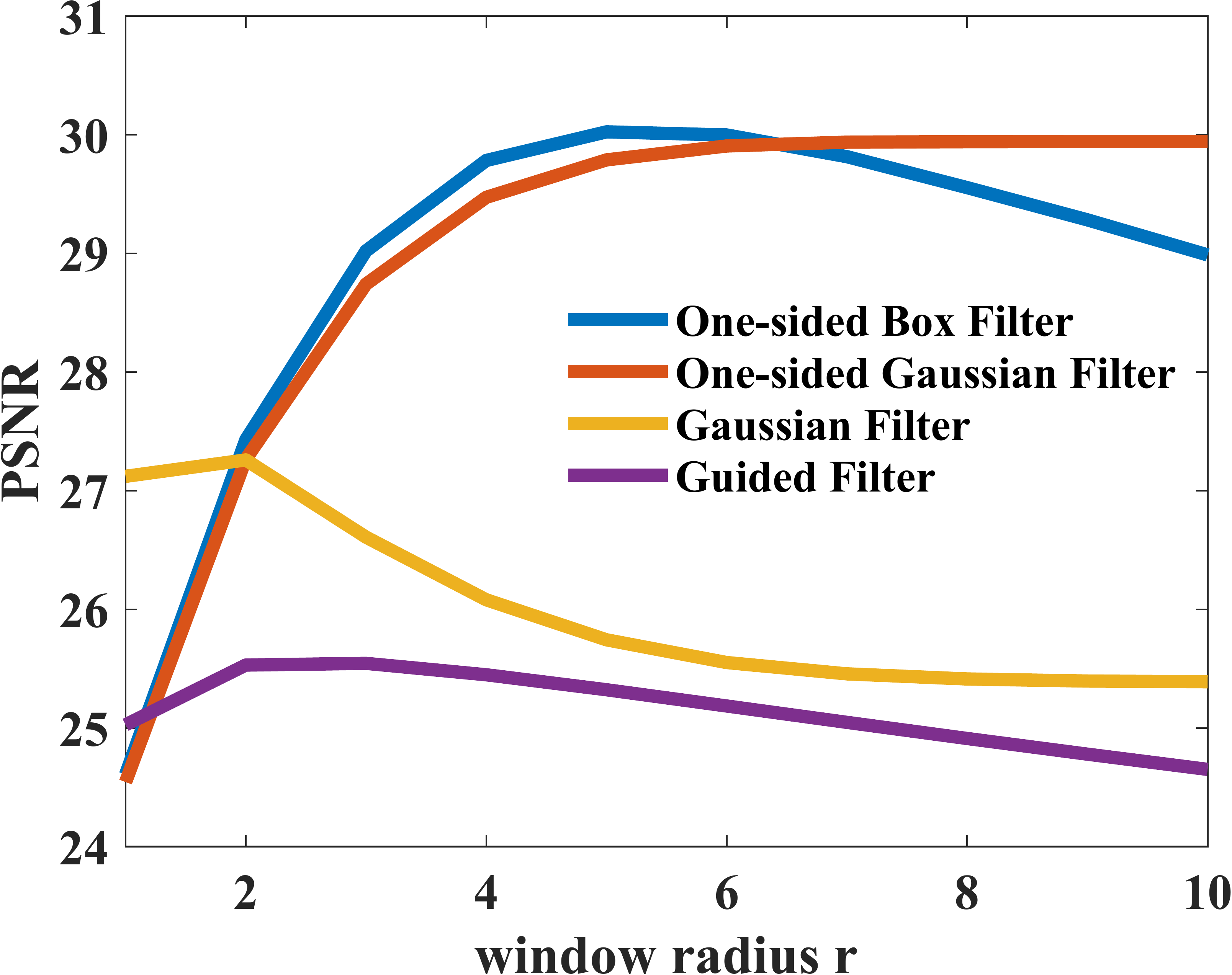}}
	\caption{Corner and Edge Preserving comparison. For the checkerboard image in (c), (d), (e) and (f), we set $r=10$.  We further increase the window radius from 1 to 10 and the PSNR for these methods is shown in (g). We fix $\sigma=3$ for Gaussian and one-sided Gaussian filters in this experiment.}
	\label{fig:checkboard}
\end{figure}

To show that our one-sided box filter preserves corners and edges, we perform it on a check board image, as shown in Fig.~\ref{fig:checkboard}. Theoretically, one-sided filters preserves corners and edges while guided filter can not preserve corners. This behavior is observed in Fig.~\ref{fig:checkboard}(g). When $r$ is small, all filters remove noise. But when $r$ further increases, Gaussian filter smooths the corners and edges. When $r$ increases, guided filter smooths out the corners. In contrast, one-sided filters preserve both corners and edges. That is the reason one-sided filters achieve higher PSNR on this synthetic image. 

In this example, one-sided Gaussian filter achieves better PSNR than the one-sided box filter when the $r\geq 7$. This phenomenon is true in general since the Gaussian weight function is spatially varying. Although the varying weights lead to better results, they also increase the computation cost.  

\subsection{Fast Approximation to One-sided Box Filter}
\label{sec:fast}
\begin{figure}[!htb]
	\centering
	\includegraphics[width=0.28\linewidth]{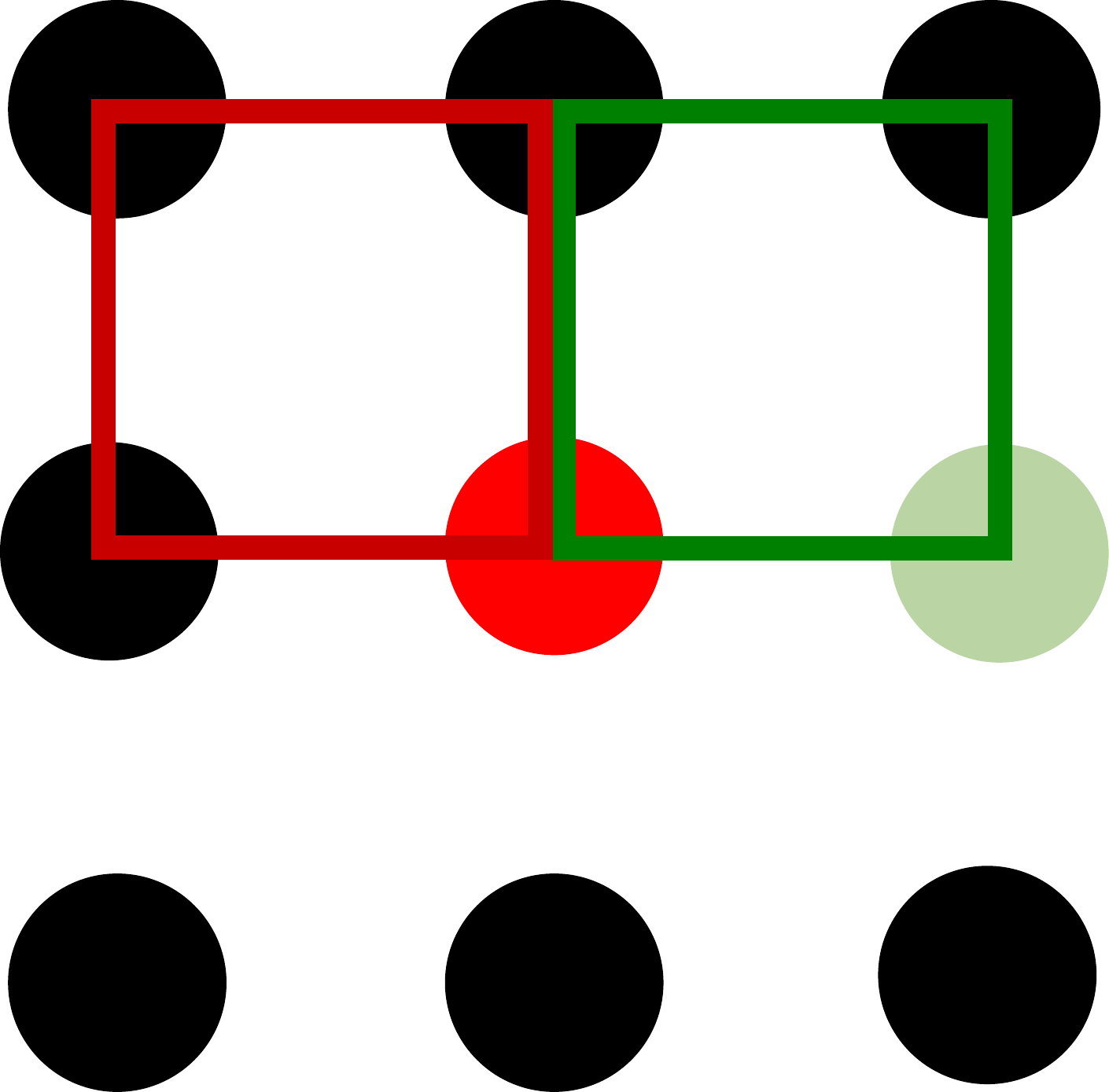}
	\caption{The right up (green) window for the red dot location (center) exactly overlaps with the left up window for the green dot location. Therefore, we only need to compute left up window average for all locations. This avoids possible redundant computation.}
	\label{fig:fast}
\end{figure} 
Although there are eight possible one-sided support regions, they have some overlapped regions. Thanks to the overlapping,  we can construct an approximation to the one-sided box filter. The approximation runs faster. And its result has no obvious visual difference from the one-sided box filter result.

As shown in Fig.~\ref{fig:fast}, the right up region of the red center location is exactly the left up region of the green location. Therefore, the potential repeated computation can be avoided. We only compute the left up region for each location. When we need other regions, for example the right up region for a location, we only need to lookup the results by shifting the location with $r$. This strategy reduces redundant computation. 

We can also use such left up region average to approximate the average values for the half window supported regions. For example, the upper half window average in Fig.~\ref{fig:fast} can be approximated by the average from the two average values based on the quarter sub windows. More specifically, let $a_1$ and $a_2$ denote the average values from ${\cal N}_1$ and ${\cal N}_2$ respectively, then the value $a_7$ from ${\cal N}_7$ can be approximated by
\begin{equation}
\label{eq:app}
	a_7\approx \hat{a}_7\equiv\frac{a_1+a_2}{2}\,.
\end{equation} The $a_5,a_6,a_8$ can be approximated in a similar way.

This equation is a numerical approximation. The reason comes from the vertical overlapped line, which is covered by both the left and right sub windows, as shown in Fig.~\ref{fig:fast}. Therefore, the samples on the vertical and horizontal line have higher weight than others.

Such higher weight is reasonable because the horizontal and vertical gradients are indeed dominants for natural images~\cite{gong:gdp}. Since the horizontal and vertical gradients have higher probability than other directional gradients, giving them higher weight is a reasonable choice from Bayesian point of view.

The fast one-sided box filter is shown in Algorithm~\ref{algo:HWB2}. It only requires one convolution during each iteration while the original Algorithm~\ref{algo:HWB} requires eight convolutions. 
\begin{algorithm}[!htb]
	\caption{Fast One-Sided Box Filter}
	\label{algo:HWB2}
	\begin{algorithmic}
		\Require $I(x,y)$, window radius $r$, IterationNum
		\State $t=0$, $U^0(x,y)=I(x,y)$
		\While{$t \leq$ IterationNum}
		
		\State $d=k_1\otimes U^{t}$ 
		\State $(u_m,v_m)=\arg\min\limits_{u,v\in\{0,r\}\}}\{|d(x+u,y+v)-U^t(x,y)|,\,|\frac{d(x+u_1,y+v_1)+d(x+u_2,y+v_2)}{2}-U^t(x,y)|\}$
		\State $U^{t+1}(x,y) = U^{t}(x,y) + d(x+u_m,y+v_m)$
		\State $t=t+1$
		\EndWhile
	\end{algorithmic}
\end{algorithm}

\begin{figure}[!htb]
	\centering
	\includegraphics[width=0.6\linewidth]{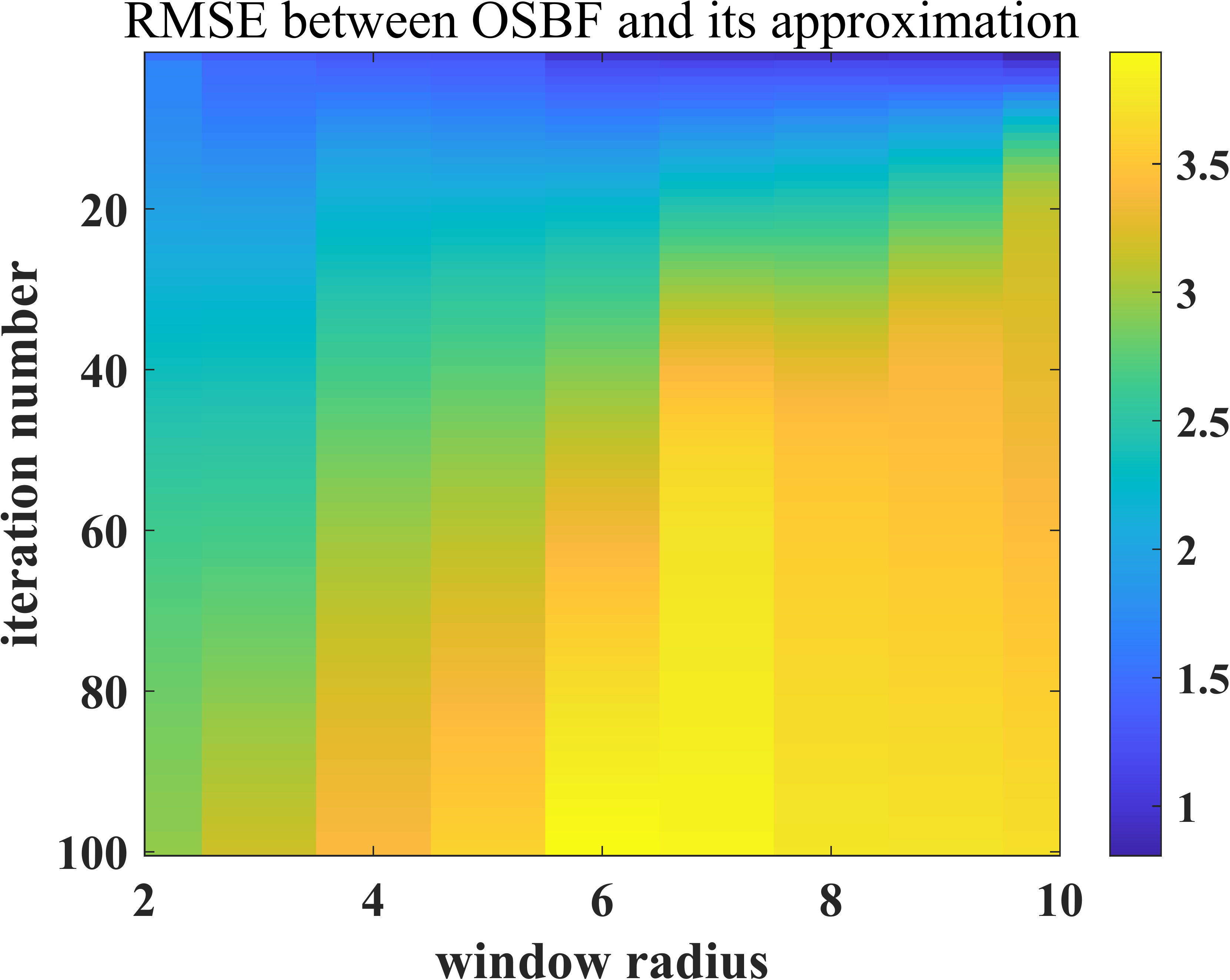}
	\caption{The average difference between one-sided box filter (Algorithm~\ref{algo:HWB}) and fast one-sided box filter (Algorithm~\ref{algo:HWB2}) on the cameraman image. The radius $r$ is increased from 2 to 10 while the iteration number is increased from 1 to 100. The max difference between these tow algorithms is shown in Fig.~\ref{fig:fastpng}. There is no obivous visual difference.}
	\label{fig:fastapp}
\end{figure}

\begin{figure}[!htb]
	\includegraphics[width=\linewidth]{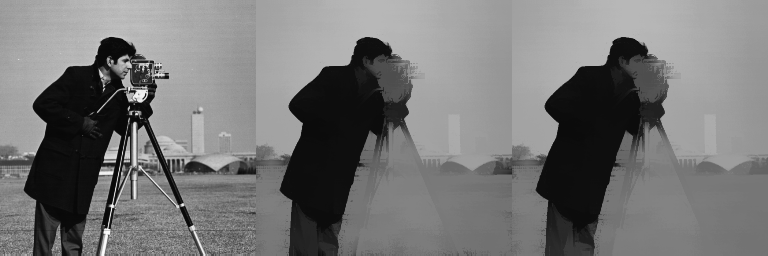}
	\caption{original (left), Algorithm~\ref{algo:HWB}: one-sided box filter with $r=6$ and 100 iterations (middle), Algorithm~\ref{algo:HWB2}: fast one-sided box filter with $r=6$ and 100 iterations (right). There is no obvious difference between the middle and right since the RMSE is about 4. But Algorithm~\ref{algo:HWB2} runs faster.}
	\label{fig:fastpng}
\end{figure}

To evaluate the approximation error, we perform both algorithms on the cameraman image with window radius increasing from 2 to 10 and iteration number increasing from 1 to 100. More specifically, let 
\begin{equation}
	F=\mathrm{OSBF}(im,r,iteration)\,,
\end{equation}
\begin{equation}
F_{fast}=\mathrm{FastOSBF}(im,r,iteration)\,.
\end{equation}The Root Mean Square Error (RMSE) between $F$ and $F_{fast}$ is shown in Fig.~\ref{fig:fastapp}. The maximum RMSE is about 4, which is too small to lead obvious visual differences. To visually confirm this, we show the results with $r=6$ and iteration=100 in Fig.~\ref{fig:fastpng} because these parameters lead to the largest RMSE. They have no obvious visual difference.

\begin{table}
	\centering  
	\begin{tabular}{c c c c c} 
		\hline
		Resolution & 256$\times$256 & 512$\times$512 & 1024$\times$1024 & 2048$\times$2048\\
		\hline
		Box Filter & 0.0007&    0.0030&    0.0106&    0.0450\\
		\hline
		OSBF(Algo.~\ref{algo:HWB}) & 0.0053 & 0.0228 & 0.0840 & 0.3597\\
		\hline
		FastOSBF(Algo.~\ref{algo:HWB2}) & 0.0024 & 0.0146 & 0.0595 & 0.2401\\
		\hline 
	\end{tabular}
	\caption{running time in seconds for these filters with one iteration. All filters are implemented by MATLAB2020a and performed on Thinkpad P1 with E-2176M CPU.} 
	\label{table:time} 
\end{table}  

\subsection{Running Time}
Thanks to the $O(1)$ and $O(N)$ complexity, the proposed filter is computationally fast. We use the classical box filter as base line. We perform box filter, one-sided box filter and its fast approximation on images with different resolutions. The running time of Algorithm~\ref{algo:HWB} and \ref{algo:HWB2} is shown in Table~\ref{table:time}, compared with the classical box filter. The running time of one sided box filter is about eight times of counterpart for box filter while both are implemented in Matlab. 

We also notice that the approximation algorithm can save about $40\%$ of the running time in practice. The hardware is Thinkpad P1 with E-2176M CPU and the software is MATLAB 2020a.

We further compared the proposed one-sided box filter with the classical box filter with C++ implementation, based on OpenCV package. The running time of OSBF is only about six times of the counterpart for box filter. Therefore, it can achieve real time performance for most applications in practice.
\subsection{Higher Dimension Case}
\label{sec:high}
\begin{figure}[!htb]
	\centering
	\includegraphics[width=0.3\linewidth]{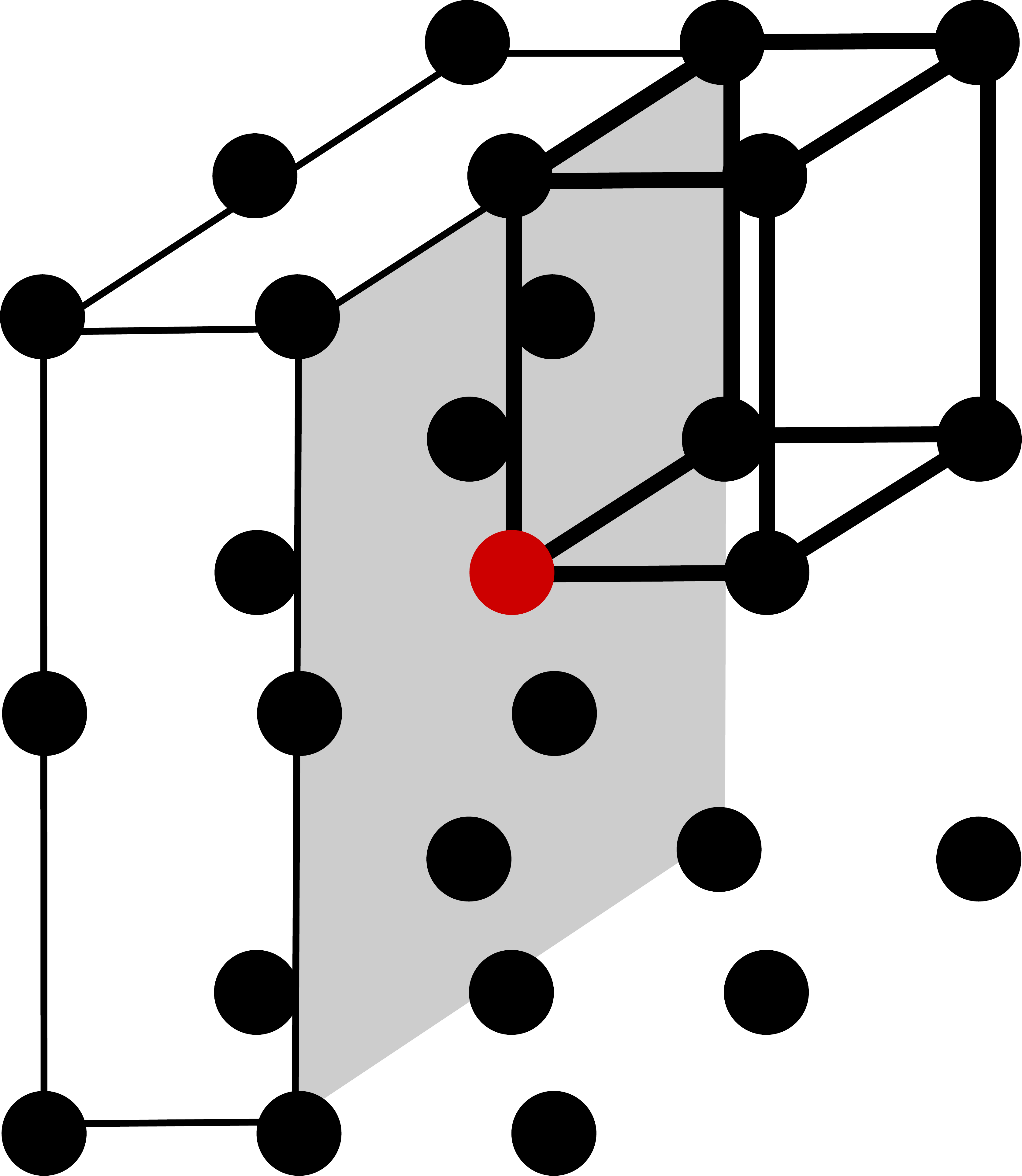}
	\caption{One half window and one corner-like sub window in 3D. There are six half windows. And there are eight such corner-like sub windows.}
	\label{fig:3D}
\end{figure}

\begin{figure}[!b]
	\centering
	\subfigure[input 3D data ]{\includegraphics[width=0.45\linewidth]{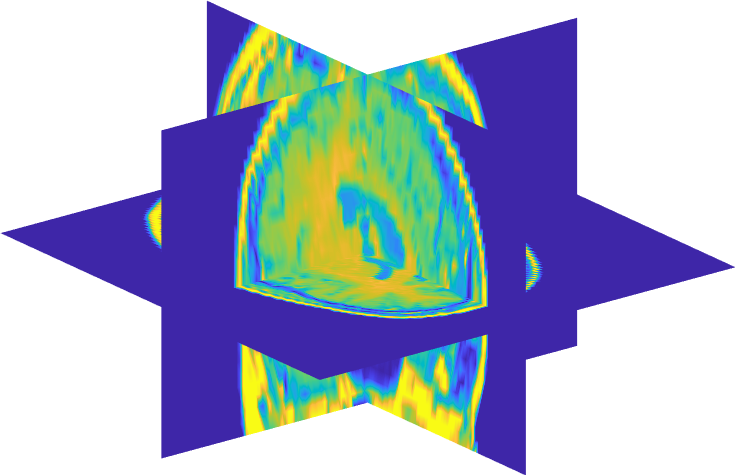}}
	
	\subfigure[result from box filter]{\includegraphics[width=0.45\linewidth]{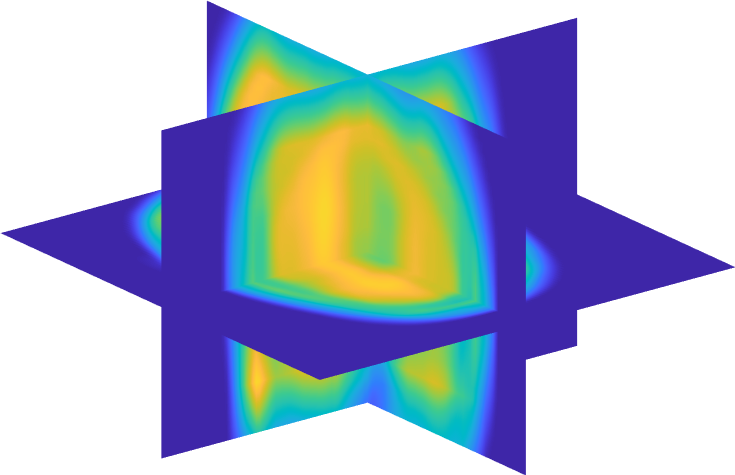}}
	\subfigure[result from one-sided box filter]{\includegraphics[width=0.45\linewidth]{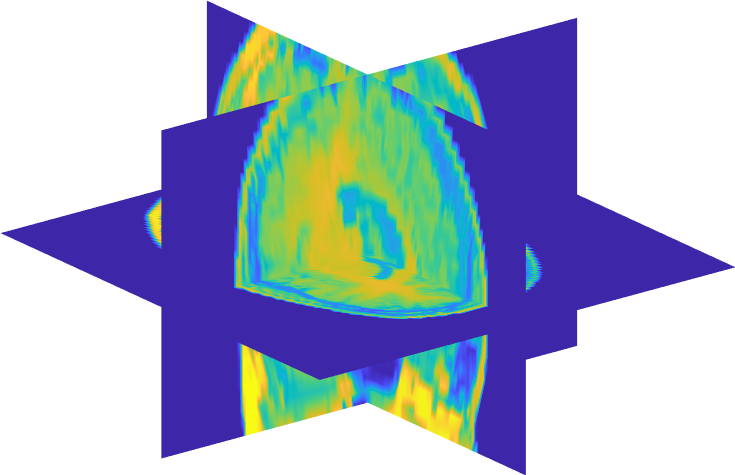}}
	\subfigure[one slice from input (left), box filter result (middle) and one-sided box filter result (right). One-sided box filter preserves edges during smoothing.]{\includegraphics[width=\linewidth]{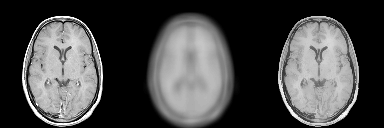}}
	\caption{Compare the one-sided box filter with the classical box filter on 3D MRI data. (a) input 3D data (only show three slices for better visualization), (b) smoothed result by box filter with $r=2$, RMSE=15.32, (c) smoothed result by one-sided box filter with $r=2$, RMSE=5.13. (e) one slice from (a), (b) and (c), respectively.}
	\label{fig:3dmri}
\end{figure}

The one-sided filter also works in higher dimensions, for example, 3D and 4D. One sub window for the 3D one-sided filter is shown in Fig.~\ref{fig:3D}. There are eight such sub windows. For these corner-like regions, we can perform following separable kernel along each dimension
\begin{equation}
	k_{x}^-=
	\begin{cases}
	\frac{1}{r+1}, &-r\leq {x}\leq 0 \cr 0, & 0< {x}\leq r
	\end{cases},\,k_{x}^+=
	\begin{cases}
	0, &-r\leq {x}< 0 \cr \frac{1}{r+1}, & 0\leq {x}\leq r
	\end{cases}.
\end{equation}

Similarly, combining these half kernels with the original box filter kernels
\begin{equation}
k_{x}=\frac{1}{2r+1}, -r\leq {x}\leq r\,
\end{equation} leads to the half window support regions. There are six such half windows in the 3D case (left, right, front, back, up and down half windows).

\begin{table}[!t]
	\centering  
	\begin{tabular}{c|cc}
		\hline
		& RMSE & running time (seconds)\\
		\hline
		box filter& 15.32 & 0.05\\
		one-sided box filter & 5.13 & 0.1\\
		\hline
	\end{tabular}
	\caption{Compare one-sided box filter with the classical box filter}\label{table:88}
\end{table}

We compare the one-sided box filter and original box filter on a 3D MRI data. For both methods, we set the window radius $r=2$ and iteration number as one. The results are shown in Fig.~\ref{fig:3dmri}. The RMSE between smoothed result and the input for box filter is 15.32. In contrast, RMSE for our method is 5.13. Such difference is visually shown in Fig.~\ref{fig:3dmri}(d). The running time for Fig.~\ref{fig:3dmri}(b) is 0.05 second while the running time for Fig.~\ref{fig:3dmri}(c) is 0.1 second. Both are implemented in MATLAB and performed on Thinkpad P1 with E-2176M CPU.

\section{Comparison with Edge Preserving Methods}
There are various image smoothing approaches that can preserve edges. In this section, we compare our filter with such methods in terms of edge preserving and running time

We compare our filter with other edge preserving filters, as shown in Fig.~18. For these color images, we perform one-sided box filter on each color channel separately. In this experiment, we set the window radius as 2 and the iteration number as 10. The parameters for other methods are the same as that in their original papers.   

\begin{figure}[!htb]
	\centering
	\includegraphics[width=0.6\linewidth]{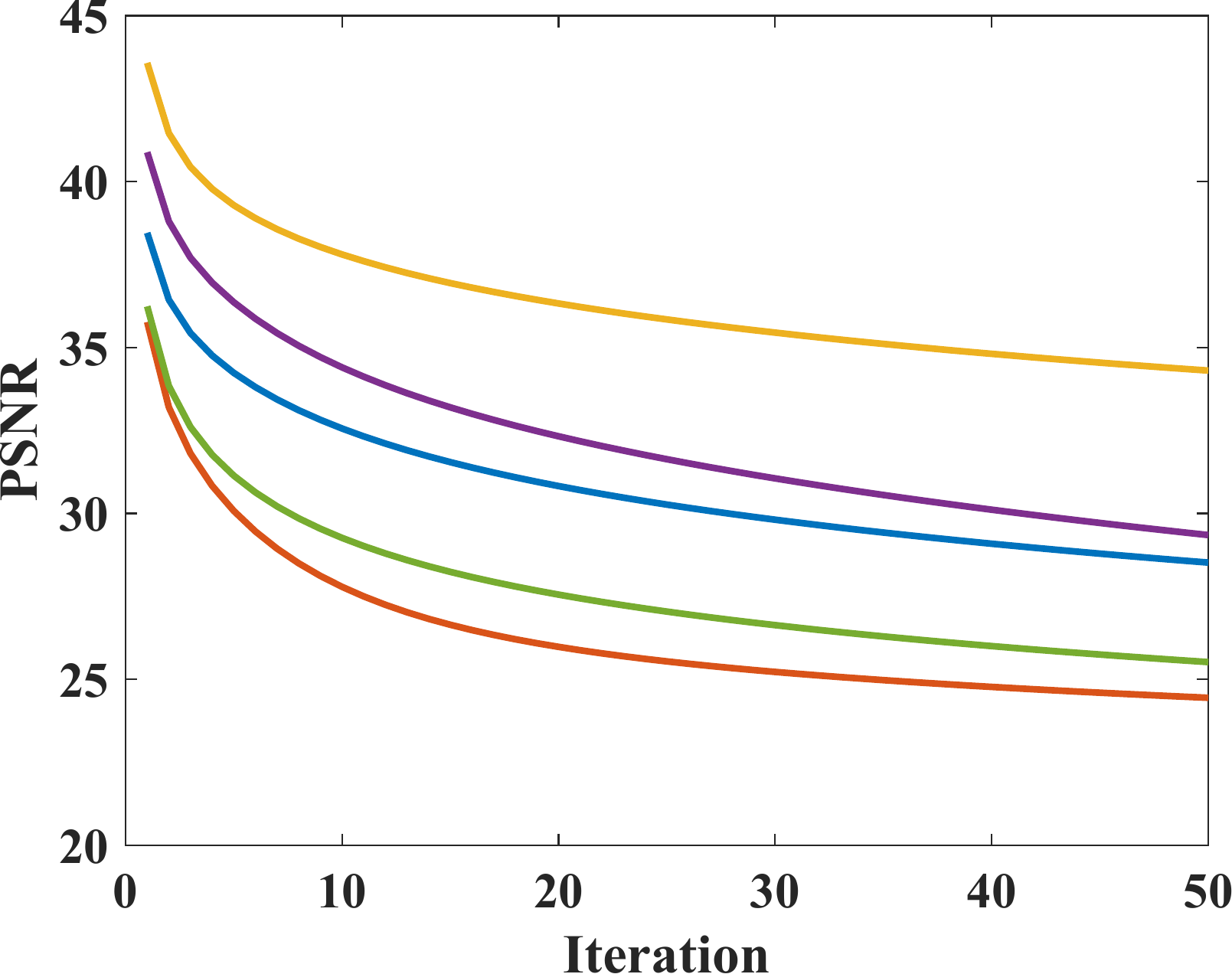}
	\caption{PSNR with different iteration numbers on different images. The line color indicates different images in Fig.~18. After ten iterations, the PSNR does not reduce significantly.}
	\label{fig:itera}
\end{figure}

Increasing the iteration number in our filter would remove more details of the input image. We increase the iteration number from 1 to 50 and perform our filter on five images in Fig.~18. The PSNR between input and output from our filter is shown in Fig.~\ref{fig:itera}. After ten iterations, the PSNR does not reduce significantly. Therefore, we choose 10 as our default iteration number.

To quantitatively show the edge preserving property, we compute the structural similarity (SSIM) between the output and input. The results are shown in Table~\ref{table:2}. Our filter can keep more structural information among these methods. The running time for these methods is shown in Table~\ref{table:3}.
\begin{table}[!t]
	\centering  
	\begin{tabular}{cccccc}
		\hline\hline
		image index &\cite{DomainTransform}&\cite{he2010guided}&\cite{Xu2012}&\cite{Ham2018} & ours\\
		1 & 0.874 & 0.953 & 0.852 & 0.867 & \textbf{ 0.976} \\
		2 & 0.900 & \textbf{ 0.990} & 0.919 & 0.973 & 0.974 \\
		3 & 0.917 & 0.986 & 0.917 & 0.925 & \textbf{ 0.993} \\
		4 & 0.863 & 0.934 & 0.863 & 0.813 & \textbf{ 0.990} \\
		5 & 0.902 & 0.952 & 0.861 & 0.885 & \textbf{ 0.965} \\
		\hline
		mean&0.891 & 0.963 & 0.878 & 0.882 & \textbf{ 0.980} \\
		\hline
	\end{tabular}
	\caption{Structural SIMilarity (SSIM) between input and output in Fig.~18. The average is shown at the bottom.}\label{table:2}
\end{table}

\begin{table}[!t]
	\centering  
	\begin{tabular}{cccccc}
		\hline\hline
		method &\cite{DomainTransform}&\cite{he2010guided}&\cite{Xu2012}&\cite{Ham2018} & ours\\
		\hline
		mean&52.97 & 0.3 & 7.309 & 44.27 & \textbf{0.0625} \\
		\hline
	\end{tabular}
	\caption{Running time in seconds on one million pixels.}\label{table:3}
\end{table}

\section{Conclusion}
In this paper, we propose to perform regression on eight one-sided support regions instead of the full local window. Such enumerable support regions significantly reduce the computation cost. For all one-sided regions, the central pixel is assumed on an edge or corner. Each one-sided support region corresponds to a possible normal direction. Therefore, the filter can NOT diffuse along that direction, leading to edge and corner preserving. This idea is fundamental and also generic for arbitrary weight functions. 

We combine one-sided support regions with the classical box filter, leading to the one-sided box filter.  The presented one-sided box filer can preserve edges and corners, which is theoretically guaranteed and numerically confirmed in this paper. The edge preserving property is comparable with other edge preserving methods.

Meanwhile, the proposed one-sided box filer is computationally fast. It has $O(1)$ computational complexity with respect to the window radius. It also has linear computational complexity $O(N)$ with respect to the total number of samples. The running time from its naive implementation is about eight times of running time from the classical box filter. Its fast approximation uses less running time but can obtain similar results.

Thanks to the widely influence of box filter, the one-sided box filter can be used in a large applications of signal processing, such as audio signal smoothing, radio frequency signal processing, image smoothing, 3D biomedical image denoising, 4D dynamic volume data processing, image registration, optical flow estimation, etc.
\if false
\section*{Acknowledgment}
This work was supported in part by the National Natural Science Foundation of China under Grant 61907031.

\fi
\appendix
\begin{lstlisting}
function result = osbf(input, radius, ItNum)
%%% one-sided box filter
r = radius;
input = single(input); 
result = input;
k = ones(2*r+1,1)/(2*r+1); %constant kernel
k_L=k;
k_L(r+2:end)=0; 
k_L = k_L/sum(k_L); 
k_R=flipud(k_L); %one-sided kernels

m = size(input,1)+2*r; 
n = size(input,2)+2*r;
offset = reshape(-m*n+1:0,m,n);
d = zeros(m,n,8,'single'); 
for i=1:size(input,3)
    im = padarray(input(:,:,i),[r,r],'replicate'); 
    for it = 1:ItNum
        d(:,:,1) = conv2(k, k_L, im,'same') - im; 
        d(:,:,2) = conv2(k, k_R, im,'same') - im;
        d(:,:,3) = conv2(k_L, k, im,'same') - im; 
        d(:,:,4) = conv2(k_R, k, im,'same') - im;
        d(:,:,5) = conv2(k_L, k_L, im,'same') - im; 
        d(:,:,6) = conv2(k_L, k_R, im,'same') - im;
        d(:,:,7) = conv2(k_R, k_L, im,'same') - im; 
        d(:,:,8) = conv2(k_R, k_R, im,'same') - im;
        %find the minimal signed distance
        tmp = abs(d); 
        [~,ind] = min(tmp,[],3); 
        index = m*n*ind + offset;
        dm = d(index);
        im = im + dm; 
    end
    result(:,:,i) = im(r+1:end-r,r+1:end-r);
end
\end{lstlisting}
\includepdf[pages={1}]{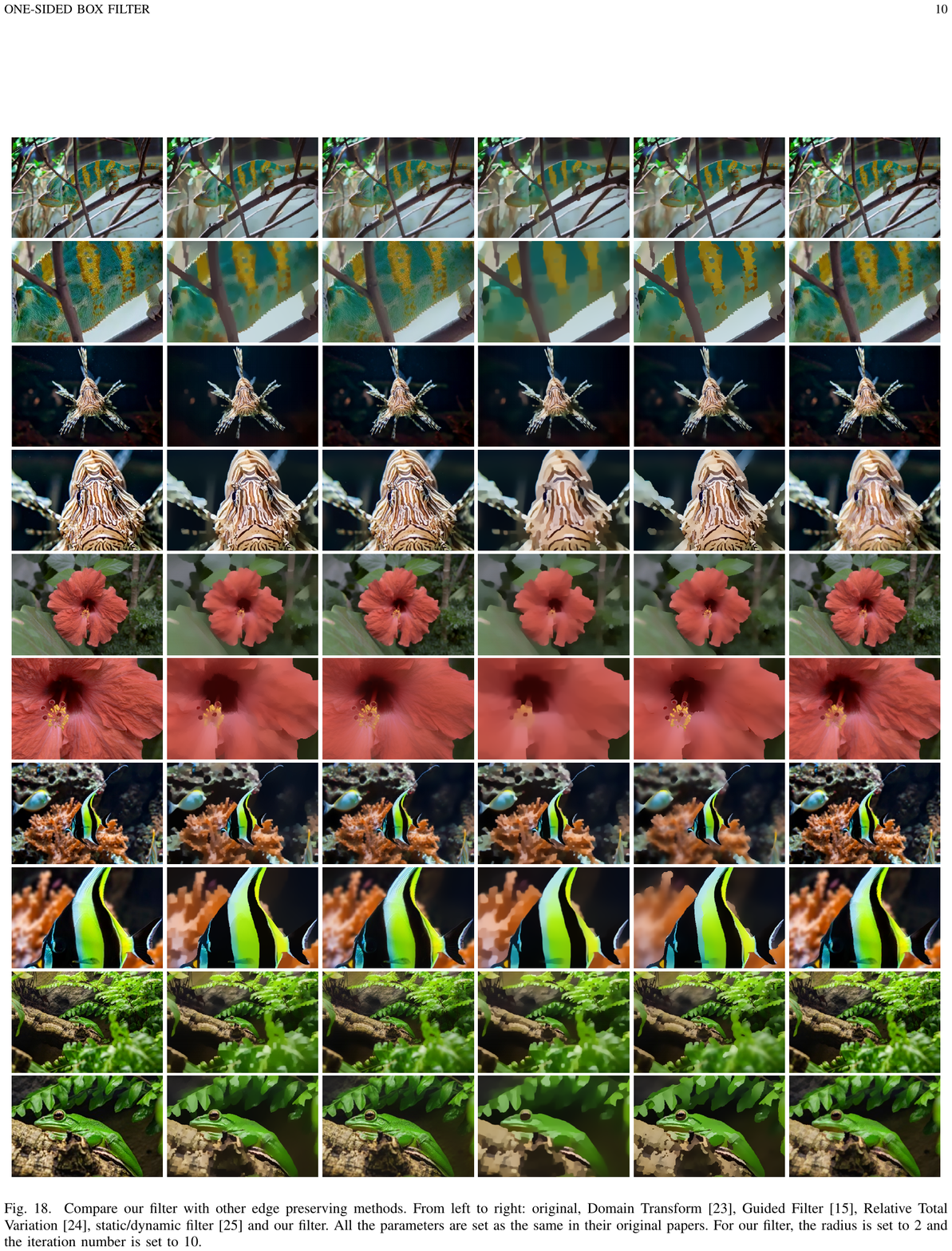}

%



%
\bibliographystyle{IEEEtran}
\bibliography{IEEEabrv,IP}

%




\end{document}